\documentclass[preprint, tighten, twocolumn]{aastex62}

\usepackage[clockwise,figuresright]{rotating}

\usepackage{amsfonts,amsmath,amssymb}

\shorttitle{The outflows driven by magnetic field and radiation force from the thin disk corona}
\shortauthors{Yang et al.}

\begin{document}

\title{Magnetohydrodynamic numerical simulation of the outflows driven by magnetic field and radiation force from the corona above a thin disk}

\author[0000-0002-2419-9590]{Xiao-Hong Yang}
\affiliation{Department of Physics, Chongqing University, Chongqing 400044, People's Republic of China;}

\author[0000-0002-0427-520X]{De-Fu Bu}
\affiliation{Key Laboratory for Research in Galaxies and Cosmology, Shanghai Astronomical Observatory, Chinese Academy of Sciences, 80 Nandan Road, Shanghai 200030, People's Republic of China}

\author[0000-0003-3706-5652]{Qi-Xiu Li}
\affiliation{Department of Physics, Chongqing University, Chongqing 400044, People's Republic of China;}

\correspondingauthor{Xiao-Hong Yang and De-Fu Bu}
\email{yangxh@cqu.edu.cn}
\email{dfbu@shao.ac.cn}

\begin{abstract}
A hot corona is suggested to be above the standard thin disk. The anisotropy of hard X-ray emission in radio-quiet active galactic nuclei implies that the corona is not static and probably moves outwards like winds. We perform two-dimensional magnetohydrodynamical simulations to study the outflowing corona driven by magnetic field and radiation force. In our simulations, as the initial state and the boundary condition at the disk surface, the corona temperature is set to $10^9$ K inside a 10 Schwarzschild radius ($r_{\rm s}$), while the corona temperature is set to $10^7$ K outside 10 $r_{\rm s}$. We employ a weak poloidal magnetic field as the initial magnetic field. A collimated outflow and a wide-angle ordered outflow are observed in our simulations. The collimated outflow is around the rotational axis and has a bulk velocity of $\sim$0.03--0.3 $c$ ( $c$ is speed of light) at 90 $r_{\rm s}$, while their mass outflow rate is very low. The collimated outflow is a weak jet. The wide-angle ordered outflow is distributed at the middle and high latitudes and moves outwards with a velocity of $10^2$--$10^4$ km s$^{-1}$. The outflow velocity depends on the disk luminosity. The gas around the disk surface is turbulent, especially outside of 10 $r_{\rm s}$. The other properties of outflows are discussed in detail.
\end{abstract}

\keywords{accretion, accretion disk -- black hole physics -- magnetohydrodynamics (MHD) -- methods}

\section{Introduction}
Observations have confirmed that outflows/winds exist in a lot of active galactic nuclei (AGNs) and X-ray binaries. Theoretically, different models are employed to explain different types of AGNs and different states of black hole (BH) X-ray binaries. In general, the hot accretion flows are used to explain the low-luminosity AGNs and the low/hard state of BH X-ray binaries (Narayan \& Yi 1994; 1995; Yuan 2001; 2003; Narayan \& McClintock 2008; Yuan \& Narayan 2014) and the standard thin disk model is used to understand the luminous AGNs (e.g. quasar) and the high/soft state of BH X-ray binaries (Shakura \& Sunyaev 1973). In the different accretion models, the formation mechanism and properties of outflows are different. In the hot accretion flows, the outflows are thought to be driven by the combination of gas pressure gradient, magnetic pressure gradient and centrifugal forces (Yuan et al. 2012; 2015; Bu \& Mosallanezhad 2018). It is difficult to directly detect the outflows in the hot accretion flows via the blueshifted spectral lines, because the high-temperature hot gas is fully ionized in the hot accretion flows. However, the indirect observational evidence implies that the outflows may appear in the hot accretion flows (e.g. Wang et al. 2013; Ma et al. 2019; Park et al. 2019).

For the luminous AGNs and the high/soft state of BH X-ray binaries, two important mechanisms are suggested to drive outflows from the thin disk around a BH. One is called the Blandford \& Payne mechanism (Blandford \& Payne 1982). In the model suggested by Blandford \& Payne (1982), the large-scale magnetic field lines are anchored in a rotating thin disk and then the outflows can be driven by the magneto-centrifugal force from the thin disk. The magneto-centrifugal outflows are strongly influenced by the inclination angle of the field lines at the disk surface. when the field lines are steep ($>60^{o}$ with respect to the disk surface), the outflows cannot directly launched from the cold disk. Cao \& Spruit (1994) present that the outflows from the disk can be loaded from a hot corona above the cold disk via thermal expansion.

The other important mechanism is that the gas above the thin disk is irradiated by the whole disk and then the radiation force exerted on the gas may overcome the effective barrier (including the effect of gravity and centrifugal force) to drive the outflows. The strength of radiation force depends on the ionization state of gas. For the highly ionized gas, the radiation force is attributed to Compton scattering. In the luminous AGNs and the high/soft state of BH X-ray binaries, the Compton-scattering force is an important force to drive the outflows when the effective barrier of gas is small (Icke 1980; Tajima \& Fukue 1996; 1998; Yang et al. 2018). For the weakly ionized gas, the radiation force can be mainly produced through the line absorption of ultraviolet photons (line force; Castor et al. 1975). The line force is important in the luminous AGNs because the luminous AGNs can emit strong ultraviolet radiation (Proga 2003; Nomura \& Ohsuga 2017), while the line force is inoperative in the high/soft state of BH X-ray binaries because of the absence of ultraviolet photons. For the luminous AGNs, the line force may also be absent in the innermost region (e.g. $<30$ $r_{\rm s}$) of the thin disk, because the strong X-ray radiation may cause the gas above the thin disk to be highly ionized in the innermost region.

Many observations suggest that outflows appear in the high/soft state of BH X-ray binaries and the luminous AGNs (e.g. Ueda et al. 1998; Reeves et al. 2003; Miller et al. 2008; Gofford et al. 2013; You et al. 2016). Several high-ionization absorption lines of metals, especially such as the Fe K-shell (Fe XXV and Fe XXVI) resonance lines, are detected in the candidates of BH X-ray binaries (Ueda et al. 1998; Kotani et al. 2000; Yamaoka et al. 2001; Lee et al. 2002; Miller et al. 2004; Miller et al. 2008). The absorption lines are found to be blueshifted, which is a significant sign of outflows. The blueshifted absorption lines are also observed in the spectra of AGNs (e.g. Pounds et al. 2003; Reeves et al. 2003; 2009; 2018; Tombesi et al. 2010; 2012; Gofford et al. 2013; 2015). Gofford et al. (2015) have detected the fast outflows from $\sim$20 \textit{Suzaku}-observed AGNs by the Fe XXVI absorption line. The fast outflows move outwards at a bulk velocity of more than 0.01 $c$, where $c$ is the light speed, and faster outflows tend to exist in more luminous AGNs. Gofford et al. (2015) imply that the fast outflows are located at $10^{2-4}$ Schwarzschild radius ($r_{\rm s}$) away from the center BH. A typical AGN, the luminous quasar PDS 456, has been studied many times (Reeves et al. 2009; 2018; Nardini et al. 2015; Matzeu et al. 2017). It was revealed in the latest study that PDS 456 harbors two components of the outflows with the Fe absorption lines (Reeves et al. 2018). One component is a wide-angle outflow with a velocity of 0.25 $c$ (Nardini et al. 2015), the other is a very fast outflow with a relativistic velocity of $\sim$0.46 $c$ (Reeves et al. 2018). The wide-angle outflow is likely launched from the inner accretion disk ($\sim$10--100 $r_{\rm s}$), while the relativistic outflow is likely launched from the innermost disk ($\sim$10 $r_{\rm s}$). Therefore, the fast outflows may arise from the accretion disk in AGNs and BH X-ray binaries.

In addition, hard X-ray emission in radio-quiet AGNs is measured to be anisotropic (e.g. Liu et al. 2014). The hard X-ray emission is believed to be produced by a hot corona near the BH. To explain the anisotropy of X-ray emission, an outflowing corona is required (Markoff et al. 2005; Liu et al. 2014). This implies that the corona is not static and probably moves outwards like winds.

In the innermost region of a thin disk, the radiation force due to Compton scattering and the large-scale magnetic field become two important factors to drive outflows, if the line force is absent. Yang et al. (2018) numerically simulated the outflows only by the radiation force due to Compton scattering from the corona above the thin disk. However, the magneto-centrifugal force and the gradient force of magnetic pressure are neglected. These two factors can play an important role in driving outflows from the thin disk (Fukue 2004; Cao 2014). The effect of magnetic field should be included. In this paper, we continue to study the potential mechanism of driving outflows from the disk corona using numerical simulations. We will further take into account the Compton-scattering force and the large-scale magnetic field at the same time, which have not been focused on by previous simulations.

This paper is organized as follows: Section 2 introduces our model and method; Section 3 is devoted to the analysis of simulations, and Section 4 summarizes and discusses our results.

\section{Model and Method}

\begin{table*}
\begin{center}

\caption{Summary of models}
%%Please Capitalize the First Letter of Each Notional Word in table's caption

\begin{tabular}{ccccccc}
\hline\noalign{\smallskip} \hline\noalign{\smallskip}

Run & $\beta_{0}$ & $\alpha_{0}$   & $\Gamma$    & $\dot{M}_{\rm out}$ ($n\times\dot{M}_{\rm C}$) & $P_{\text{k,out}}$ ($n\times L_{\rm Edd}$) & $P_{\text{th,out}}$ ($n\times L_{\rm Edd}$) \\

(1) & (2)      & (3)      &  (4)  & (5)   & (6) & (7)  \\
\hline\noalign{\smallskip}

Rwind1    & 0 &/ & 0.25   & 1.5$\times10^{-3}$     &1.9$\times10^{-10}$ &7.2$\times10^{-9}$ \\

Rwind2    & 0 &/ & 0.75   & 1.7$\times10^{-2}$     &5.9$\times10^{-6}$ &3.1$\times10^{-6}$ \\

RMwind1   & $10^{-2}$ &5.0  &0.25  & 2.8$\times10^{-3}$     &3.8$\times10^{-7}$ &1.3$\times10^{-6}$ \\

RMwind2   & $10^{-2}$ &5.0  &0.75  & 1.9$\times10^{-2}$     &7.9$\times10^{-6}$ &3.5$\times10^{-6}$ \\

\hline\noalign{\smallskip} \hline\noalign{\smallskip}
\end{tabular}
\end{center}

\begin{list}{}
\item\scriptsize{Column 1: model names. Columns 2 and 3: values of $\beta_0$ and $\alpha_0$, which determine the strength and inclination of initial magnetic field, respectively. Column 4: the ratio of disk luminosity to Eddington luminosity ($\Gamma=L_{\rm D}/L_{\rm Edd}$). Columns 5--7: the time-averaged values of the mass outflow rate, and the kinetic and thermal energy fluxes carried out by the outflows at the outer boundary. The letter ``$n$'' is defined in Equation (15).}
\end{list}
\label{tab1e_1}
\end{table*}

\subsection{Basic Equations}
We assume that the accretion disk around a BH is geometrically thin and optically thick (Shakura \& Sunyaev 1973). The thin disk can irradiate the corona above the disk surface. To simulate the irradiated corona, our computation is implemented in a spherical coordinate ($r$,$\theta$,$\phi$) and we axis-symmetrically solve the following magentohydrodynamical (MHD) equations:
\begin{equation}
\frac{d\rho}{dt}+\rho\nabla\cdot {\bf v}=0,
\label{cont}
\end{equation}
\begin{equation}
\rho\frac{d {\bf v}}{dt}=-\nabla P-\rho\nabla
\psi+\frac{1}{4 \pi}(\nabla \times \bf B)\times {\bf B}+\rho {\bf F}^{\rm rad},
\label{monentum}
\end{equation}
\begin{equation}
\rho\frac{d(e/\rho)}{dt}=-P\nabla\cdot {\bf v},
\label{energyequation}
\end{equation}
\begin{equation}
\frac{\partial {\bf B}}{\partial t}=\nabla\times({\bf v}\times{\bf B}).
\label{induction_equation}
\end{equation}
Here, $d/dt(\equiv \partial / \partial t+ \mathbf{v} \cdot \nabla)$ denotes the Lagrangian derivative; $\rho$ , $P$, $\mathbf{v}$, $\psi$ , $e$, and $\mathbf{B}$ are density, pressure, velocity, gravitational potential, internal energy, and magnetic field, respectively. We adopt an equation of state of ideal gas $P=(\gamma -1)e$ and set the adiabatic index $\gamma =5/3$. We also apply the pseudo-Newtonian potential, $\psi=-GM_{\rm BH}/(r-r_{\rm s})$  (Paczy\'{n}sky \& Wiita 1980), where $M_{\rm BH}$ and $G$ are the black hole mass and the gravitational constant, respectively, and $r_{\rm s} \equiv 2GM_{\rm BH}/c^2$. ${\bf F}^{\rm rad}$ is the total radiation force per unit mass exerted by the disk.

In the energy equation (i.e. Equation \ref{energyequation}), we assume that the irradiated gas is isentropic. This means that cooling can balance heating and then only the compression work can change the internal energy of gas. The corona above a thin disk can be heated by some physical mechanisms. Liu et al. (2002) suggested magnetic reconnection to heat the corona. Jiang et al. (2014) found dissipation of turbulence driven by MRI to be a possible mechanism. However, the heating mechanism of corona is not fully understood and is still an open issue. On the other hand, the hot corona can cool by Compton scattering, thermal conduction, and cycle-synchrotron radiation. For simplicity, the isentropic assumption is employed in our simulations.

We numerically solve Equations \ref{cont}--\ref{induction_equation} using the PLUTO code. PLUTO is a finite-volume, shock-capturing code designed to solve the HD and MHD equations (Mignone et al. 2007; 2012).

\subsection{Radiation Force}
We assume that a standard thin disk is around a BH (Shakura \& Sunyaev 1973). The total luminosity of the disk is $L_{_{\rm D}}=\Gamma L_{\rm Edd}=GM_{\rm BH}\dot{M}/(2r_{\star})$, where $\Gamma$ is the Eddington ratio of the disk luminosity, $\dot{M}$ is the accretion rate through the thin disk, and $r_{*}$ is the marginally stable orbit ($\equiv 3 r_{\rm s}$). The gas above the disk surface is irradiated by the whole disk. We assume that the irradiated gas is optically thin for the radiation of the disk, which is reasonable because the optical depth from infinity to the photosphere should be less than 1. If only Compton scattering is considered, we can integrate the radiation force ( ${\bf F}^{\rm rad}(r,\theta)$) exerted on the gas element at ($r,\theta$) in the following:
\begin{equation}
{\bf F}^{\rm rad}(r,\theta) = \Gamma\frac{6}{\pi} \frac{G M_{\rm BH}}{r^2_{*}}\int^\pi_0 \int^{1000r_{*}/3}_{r_{*}} \hat{\bf n}\frac{r' {\rm cos}(\theta)}{r'^2_{_{\rm D}}d'^{3}_{_{\rm D}}}{\rm d}r'_{_{\rm D}}{\rm d}\phi_{_{\rm D}}.
\label{rad_force}
\end{equation}
 Here, the primed quantities are expressed in units of $r_{*}$ . A polar coordinate ($r_{_{\rm D}}$,$\phi_{_{\rm D}}$) is used to define the location of radiation sources on the disk surface. $r_{_{\rm D}}$ is the distance measured from a radiation source on the disk surface to the BH; $\phi_{_{\rm D}}$ is the angle measured from the radiation source to the computational plane ($r,\theta,0^o$). $d_{_{\rm D}}$ and $\hat{\bf n}(\equiv(n_r,n_\theta,n_{\phi}))$ are the distance from the radiation source to the irradiated gas element at ($r,\theta$) and the unit vector directing toward the gas element, respectively. $\hat{\bf n}$ is given by
\begin{equation}
\begin{split}
&n_r=\frac{r-r_{_{\rm D}}{\rm sin}(\theta){\rm cos}(\phi_{_{\rm D}})}{d_{_{\rm D}}},n_\theta=-\frac{r_{_{\rm D}}{\rm cos}(\theta){\rm cos}(\phi_{_{\rm D}})}{d_{_{\rm D}}},\\
&{\rm and}\text{  }n_\phi=-\frac{r_{_{\rm D}}{\rm sin}(\phi_{_{\rm D}})}{d_{_{\rm D}}}.
 \end{split}
\end{equation}
Our computation is implemented under the axis-symmetrical assumption, so that we integrate the radial and angular components of radiation force using equation (\ref{rad_force}).

\begin{figure}
\scalebox{0.5}[0.5]{\rotatebox{0}{\includegraphics[bb=50 20 500 360]{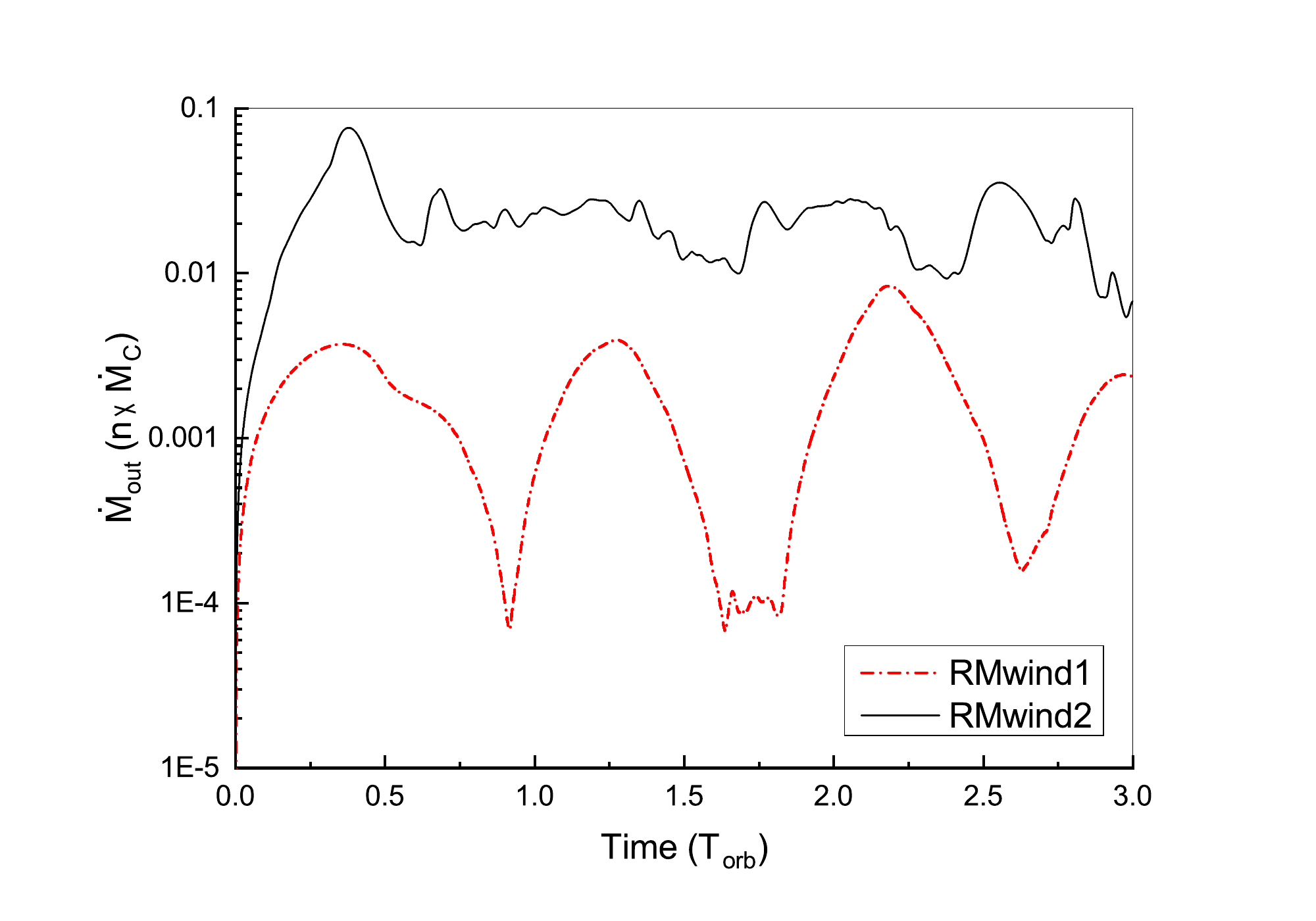}}}

\ \centering \caption{Time evolution of mass outflow rate at the outer boundary for runs RMwind1 and RMwind2. The units of vertical axis is $n\times \dot{M}_{\rm C}$, where $n$ is defined in Equation (\ref{rho_0}). }

 \label{fig_1}
\end{figure}

\begin{figure*}
\scalebox{0.45}[0.55]{\rotatebox{0}{\includegraphics[bb=80 350 520 705]{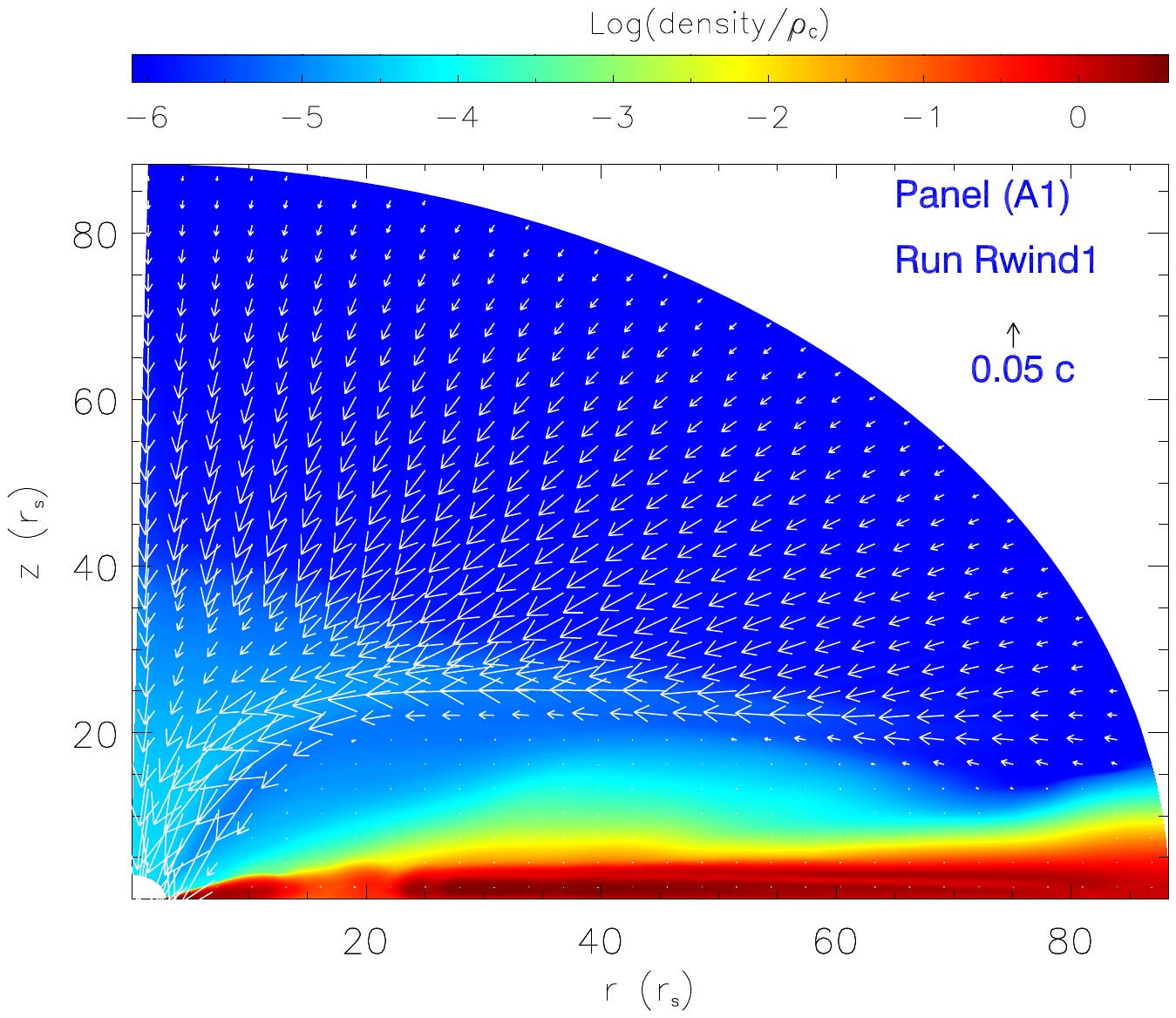}}}
\scalebox{0.45}[0.55]{\rotatebox{0}{\includegraphics[bb=80 350 520 705]{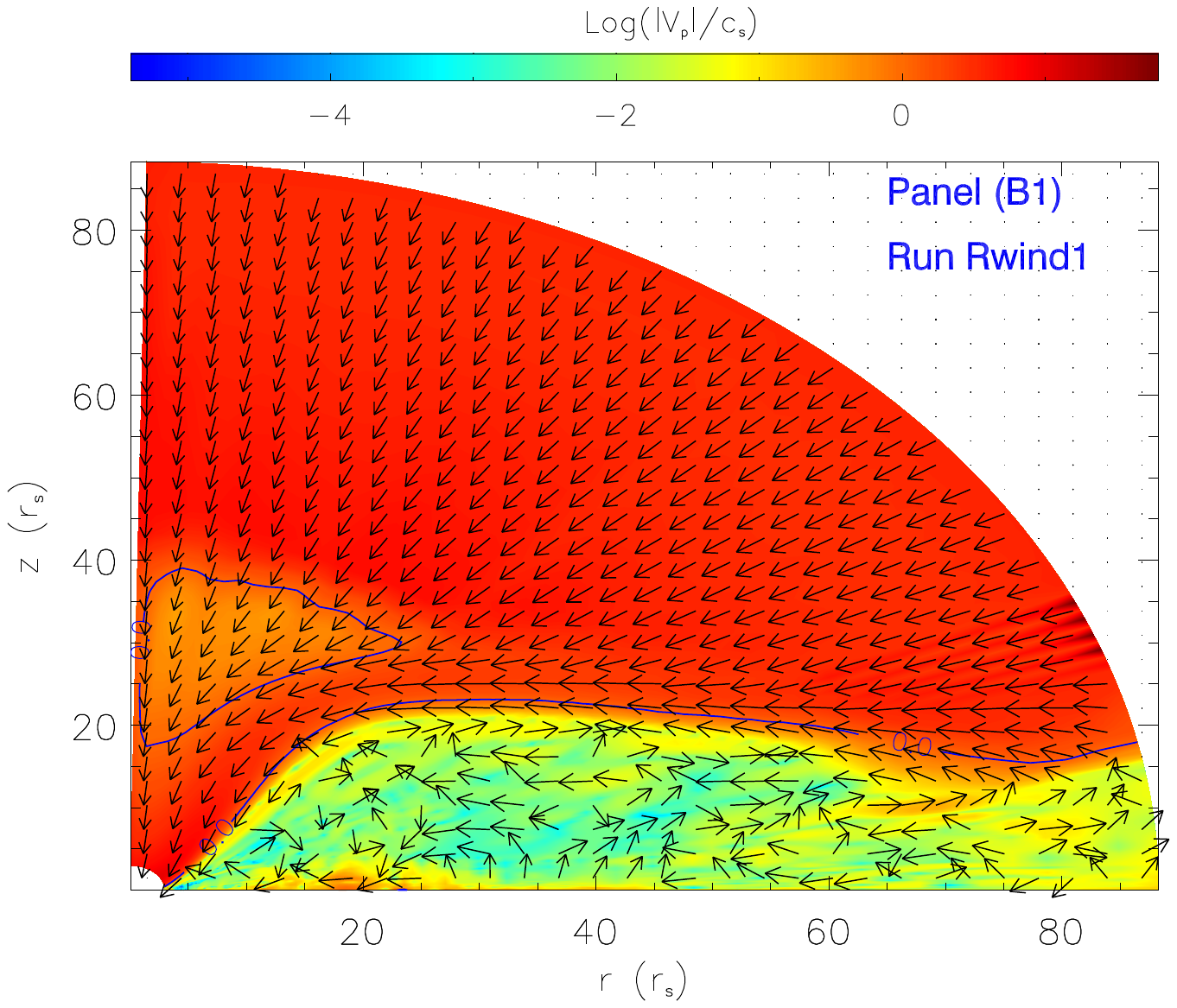}}}

\scalebox{0.45}[0.55]{\rotatebox{0}{\includegraphics[bb=80 350 520 720]{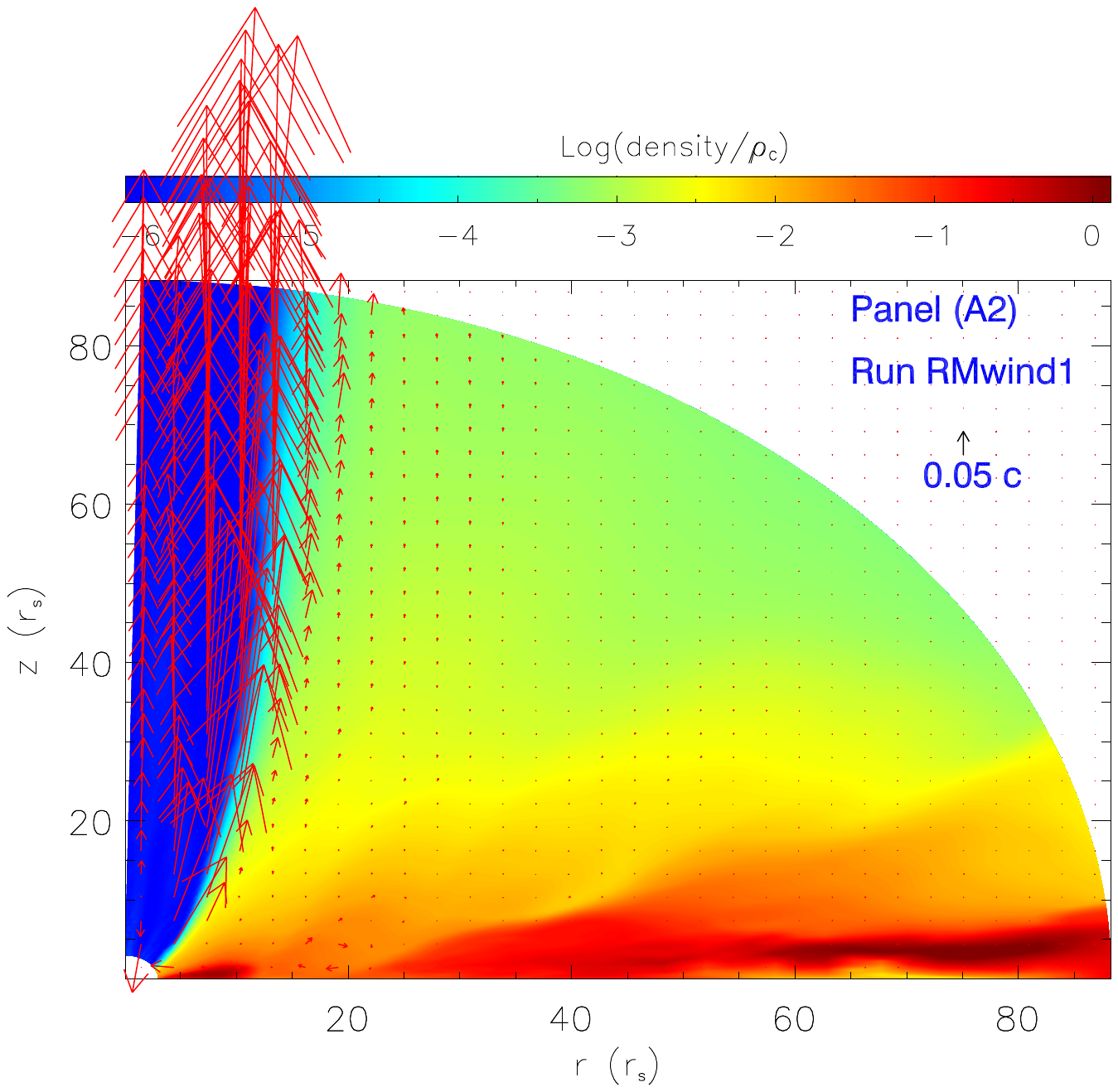}}}
\scalebox{0.45}[0.55]{\rotatebox{0}{\includegraphics[bb=80 350 520 720]{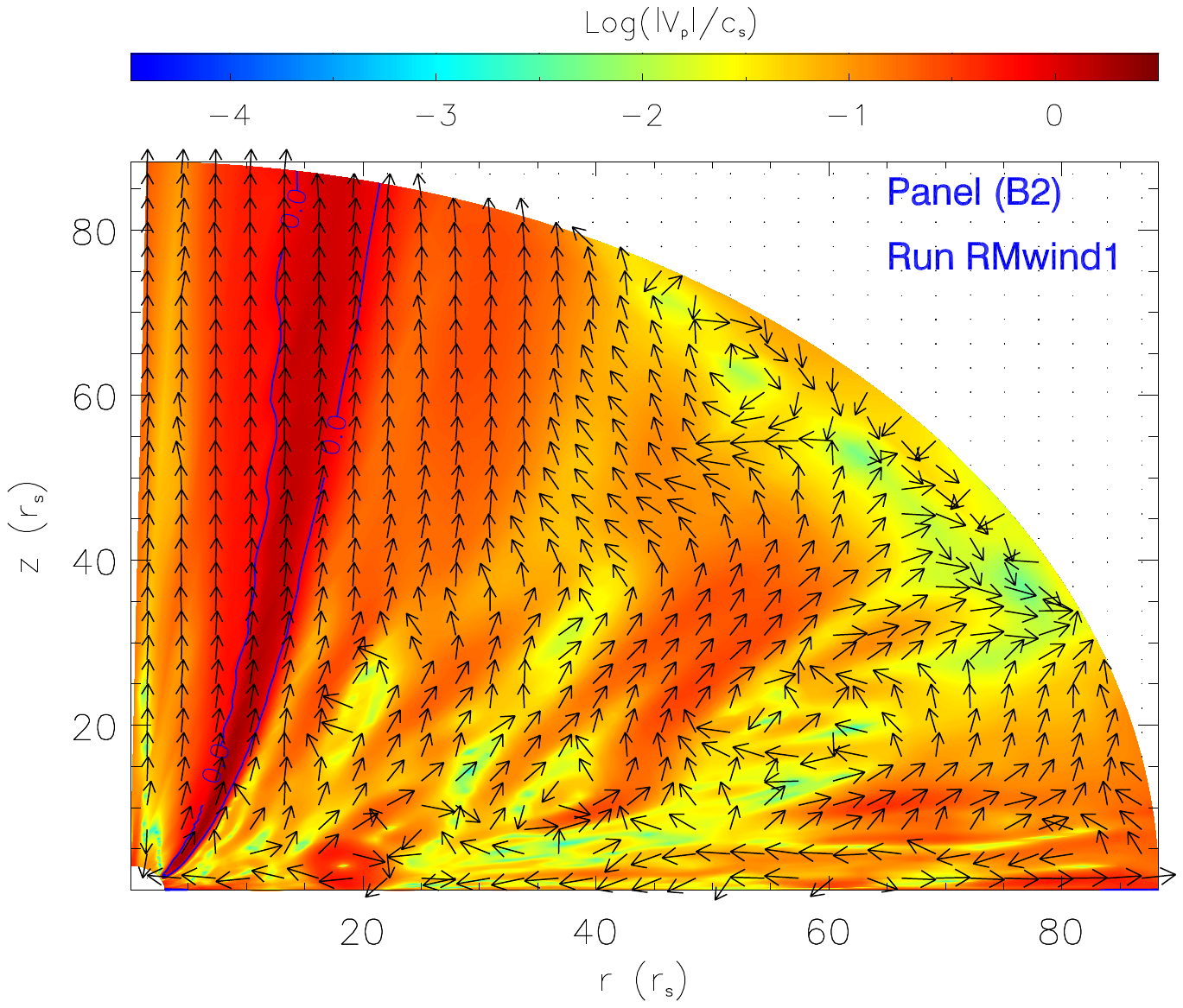}}}

\ \centering \caption{Time-averaged density, Mach number, and poloidal velocity over $t=2.0$--3.0 $T_{\rm orb}$, for runs Rwind1 (top panels) and RMwind1 (bottom panels). In panels (A1) and (A2), color means the logarithmic density, arrows mean the poloidal velocity, and the length of arrows is proportional to $|v_{\rm p}| (\equiv\sqrt{v_{r}^2+v_{\theta}^2})$. In panels (B1) and (B2), color means the logarithmic values of Mach number and arrows imply the direction of poloidal velocity. The blue solid line overplotted marks the surface of $|v_{\rm p}|=c_{s}$. The color of the arrows is just to see clearly.}

\label{fig_2}
\end{figure*}

\begin{figure*}
\scalebox{0.45}[0.55]{\rotatebox{0}{\includegraphics[bb=80 350 520 705]{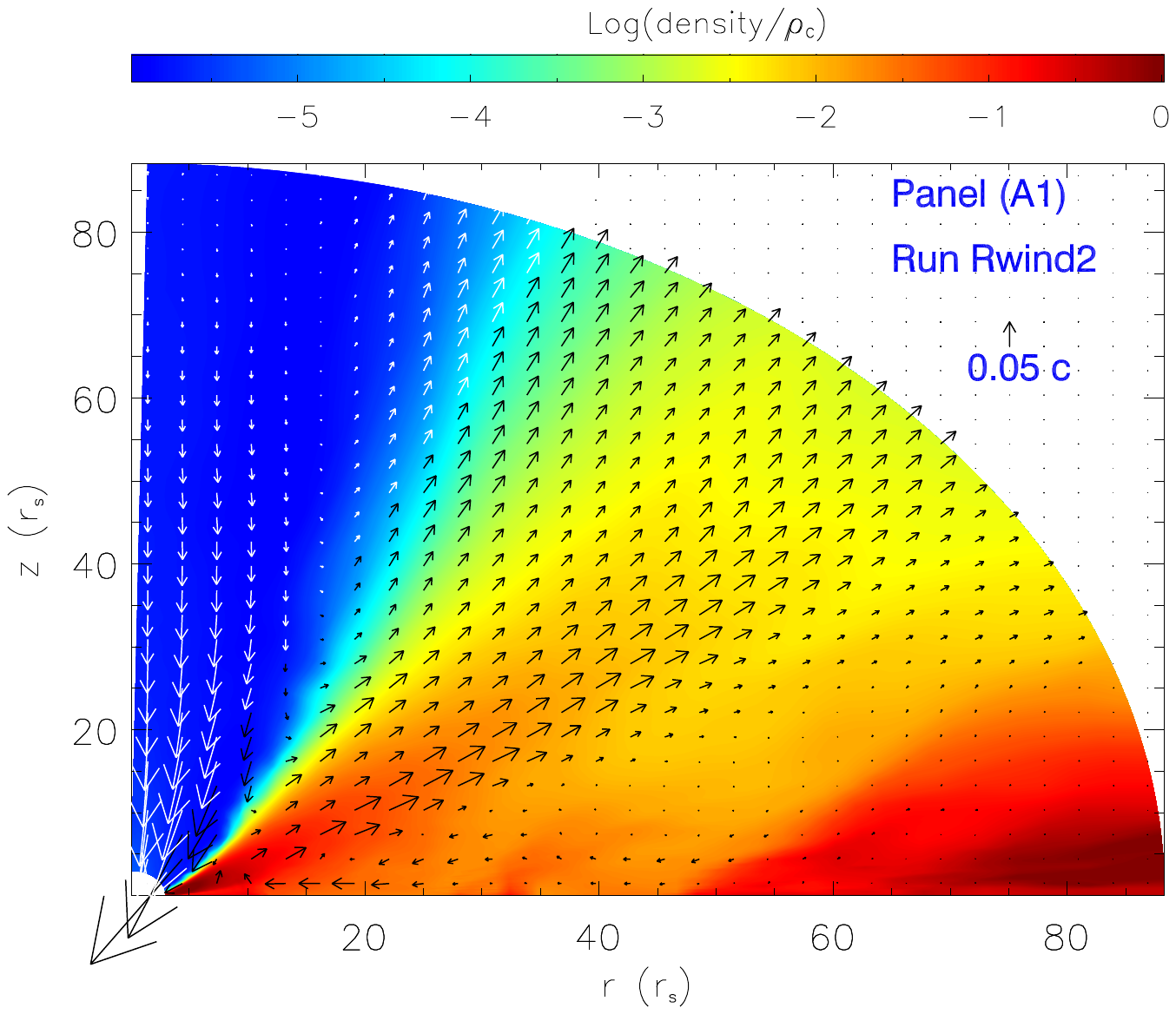}}}
\scalebox{0.45}[0.55]{\rotatebox{0}{\includegraphics[bb=80 350 520 705]{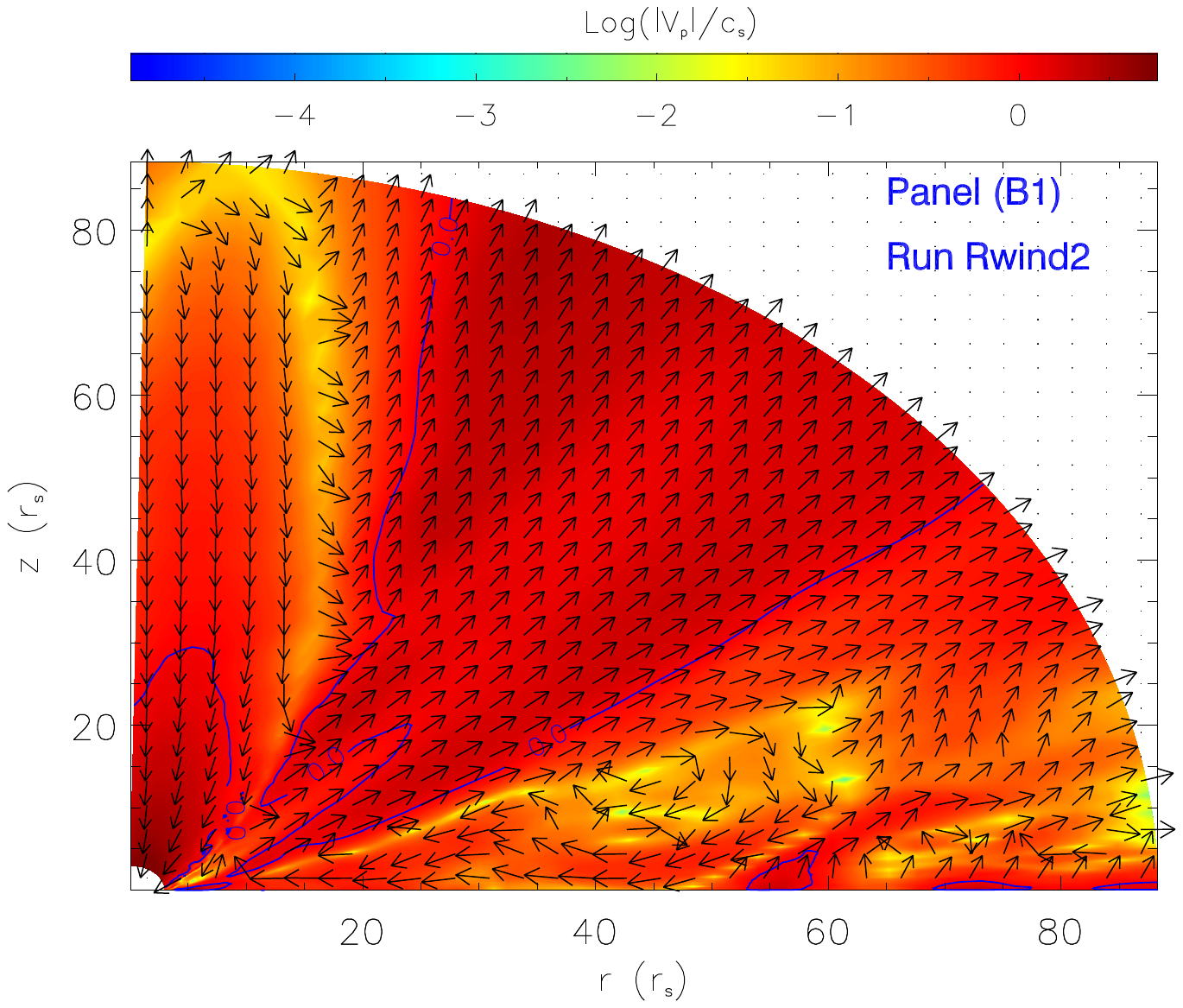}}}
%\scalebox{0.40}[0.50]{\rotatebox{0}{\includegraphics[bb=80 350 520 700]{fig3_Tg_Rwind2.ps}}}

\scalebox{0.45}[0.55]{\rotatebox{0}{\includegraphics[bb=80 350 520 700]{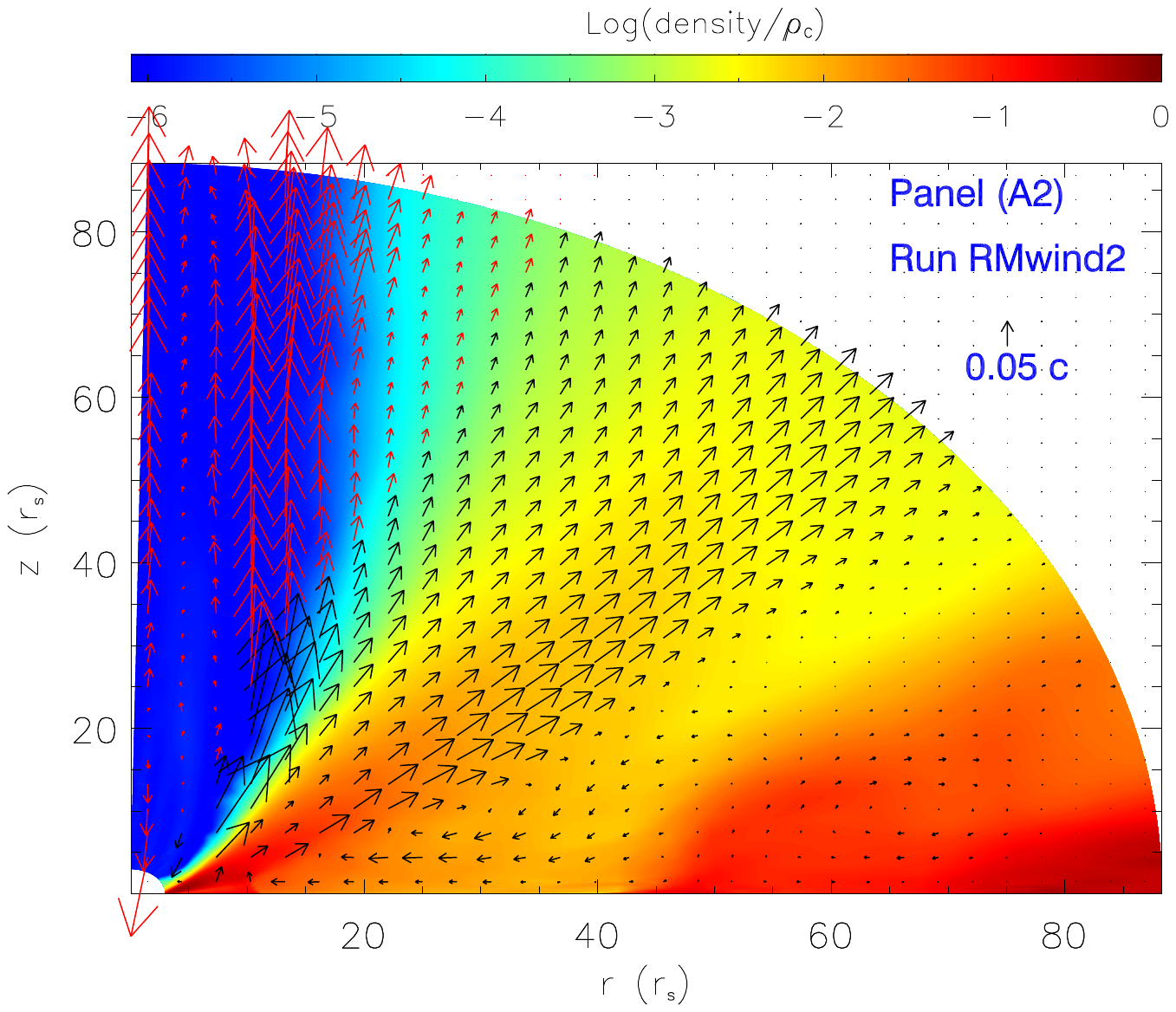}}}
\scalebox{0.45}[0.55]{\rotatebox{0}{\includegraphics[bb=80 350 520 700]{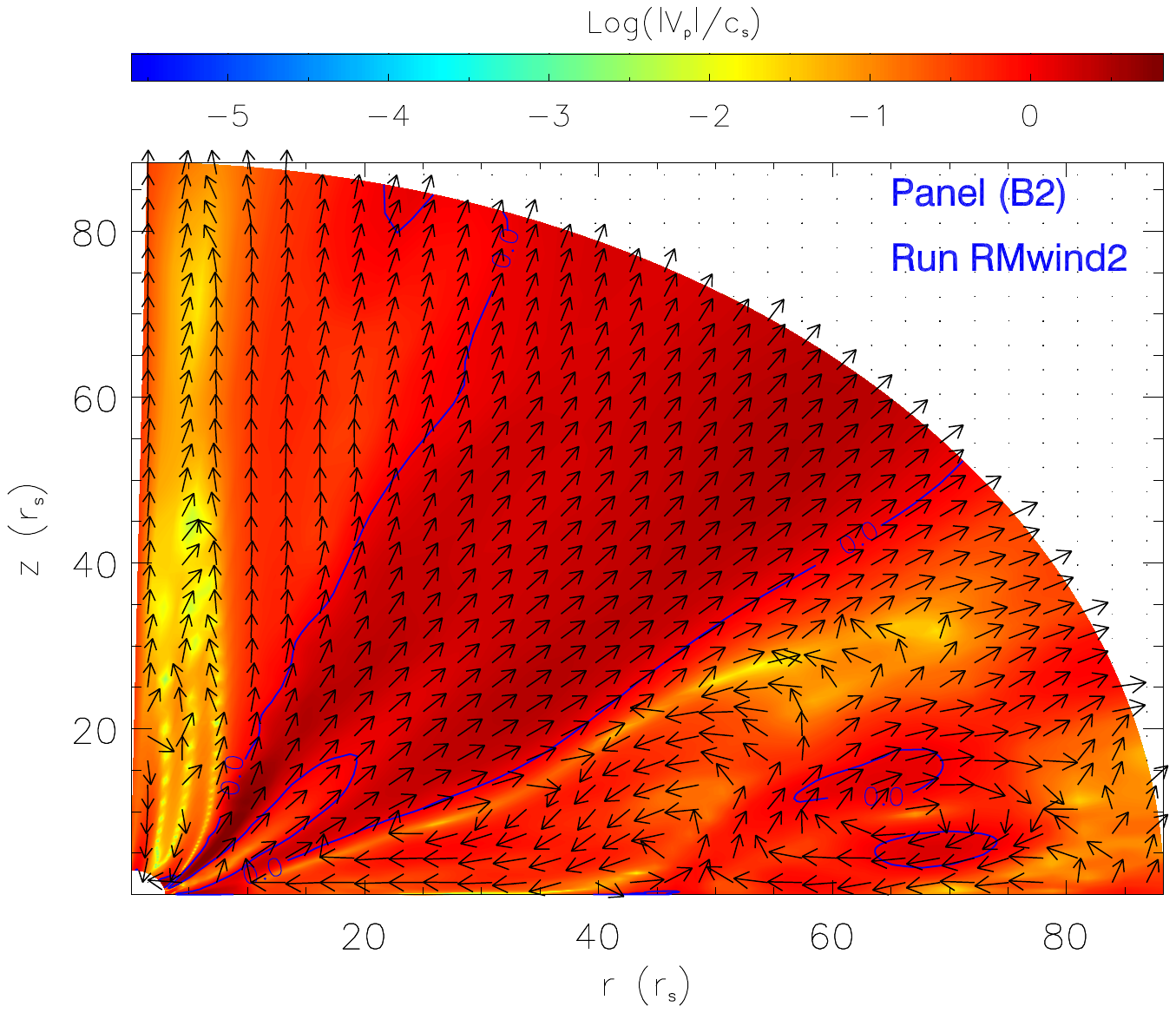}}}
%\scalebox{0.40}[0.50]{\rotatebox{0}{\includegraphics[bb=80 350 520 700]{fig3_Tg_RMwind2.ps}}}

\ \centering \caption{Same Figure \ref{fig_2}, but for runs Rwind2 (top panels) and RMwind2 (bottom panels). }

\label{fig_3}
\end{figure*}

\subsection{Magnetic Field Configuration}
The magnetic field can be written as ${\bf B}={B}_r {\bf e}_r+{B}_\theta {\bf e}_\theta+{B}_\phi {\bf e}_\phi$ in a spherical coordinate system. Under the axisymmetric assumption, the solenoidal condition $\nabla \cdot {\bf B}=0$ makes the poloidal field (${\bf B}_{\rm p}\equiv{B}_r {\bf e}_r+{B}_\theta {\bf e}_\theta$) keep $\nabla \cdot {\bf B}_{\rm p}=0$. ${B}_\phi$ becomes an arbitrary function of $r$ and $\theta$. To keep $\nabla \cdot {\bf B}_{\rm p}=0$, a vector potential ${\bf A}\equiv(0,0,{\bf A}_{\phi})^{\rm T}$ is introduced to determine the magnetic field, i.e. ${\bf B}_{\rm p}=\nabla \times {\bf A}_{\phi}$. Initially, we adopt the magnetic field configuration given in Cao \& Spruit (1994). In the spherical coordinate, the toroidal component of the vector potential is written as
\begin{equation}
\begin{aligned}
{\bf A}_\phi=\frac{\Phi_{0}}{rsin(\theta)}\frac{r_{\star}^2}{\alpha_0^2}[\sqrt{(\alpha_0\frac{r}{r_{\star}}\sin(\theta))^2+(1+\alpha_0\frac{r}{r_{\star}}\cos(\theta))^2}\\
-|1+\alpha_0\frac{r}{r_{\star}}\cos(\theta)|],
\label{vectorpotential}
\end{aligned}
\end{equation}
where $\Phi_{0}$ and $\alpha_0$ are two parameters to determine the strength and inclination of initial magnetic field, respectively. The field components are give by $B_{r}=\frac{1}{r sin(\theta)}\frac{\partial({\bf A}_\phi sin(\theta))}{\partial \theta}$ and $B_{\theta}=-\frac{1}{r}\frac{\partial({\bf A}_\phi r)}{\partial r}$. At the plane of $\theta=\pi/2$, the magnetic field configuration is given by
\begin{equation}
B_{r}(r,\theta=\pi/2)=\frac{\Phi_{0} r_{\star}}{\alpha_0 r}[1-\frac{1}{\sqrt{1+(\alpha_0 \frac{r}{r_{\star}})^2}}],
\end{equation}
\label{B_r}
and
\begin{equation}
B_{\theta}(r,\theta=\pi/2)=-\frac{\Phi_{0}}{\sqrt{1+(\alpha_0\frac{r}{r_{\star}})^2}}.
\label{B_theta}
\end{equation}
The configuration of magnetic field is similar to that used in Blandford \& Payne (1982), who studied the self-similar disk wind.

We use $\beta_{0}=B({r_{\star},\pi/2})^2/(8\pi P_{_{\rm C,}r_{\star}})$ to scale the magnitude of the magnetic field at $r=r_{\star}$. $P_{_{\rm C,}r_{\star}}$ is the gas pressure of corona above the disk surface at $r=r_{\star}$. Then, we can obtain $\Phi_{0}^2=\beta_0 P_{_{\rm C,}r_{\star}}[1+\alpha_0^2+\sqrt{1+\alpha_0^2}]$.

\subsection{Model Setup}
Our goal is to simulate the outflows launched from the corona above the disk surface. The Simulation domain, whose radial range is from $r_{*}$ to $30 r_{*}$, is located above the disk surface. The thin disk is not included in our computational domain. The disk corona is taken as the simulation boundary. For simplicity, we make no distinction between the disk surface and the disk midplane. Physical quantities at $\theta=\pi/2$ are set according to the properties of the corona. The density ($\rho_{_{\rm C}}$) of the corona is assumed to be constant at all radii. Observations have suggested that strong hard X-ray emitting regions are highly compact and are in within 10 $r_s$ (Reis \& Miller 2013; Uttley et al. 2014). The hard X-ray emission needs a hot corona whose temperature is around $10^9$ K for AGNs (e.g., Liu et al. 2003; Cao 2009).  This implies that the corona of $10^9$ K exists within 10 $r_s$. Local numerical simulations of the thin disk found that a corona of $10^7$ K can form above the disk surface via the dissipation of turbulence driven by MRI in the thin disk (Jiang et al. 2014). Therefore, we assume the temperature ($T_{_{\rm C}}$) of the corona at disk surface as follows
\begin{equation}
T_{_{\rm C}}(r)=10^{5+\frac{8}{\tanh(\frac{r}{r_{\rm s}}-12)+3}} \text{ K}.
\label{Tc}
\end{equation}
Equation \ref{Tc} implies that $T_{_{\rm C}}$ is nearly equal to $10^9$ K within 9 $r_s$, while $T_{_{\rm C}}$ is nearly equal to $10^7$ K outside 14 $r_s$. The radial and vertical velocity of the corona are assumed to be null and the corona has the same rotating velocity with the disk. Cao (2012) studied the structure of a radiation-pressure-dominated accretion disk with a large-scale magnetic field. When the effect of magnetic field is taken into account, the circular motion of the thin disk becomes sub-Keplerain. We follow Cao (2012) to calculate ${\bf v}_{\phi}(r,\pi/2)$ using
\begin{equation}
{\bf v}_{\phi}(r,\pi/2)^2=(r\Omega_{\rm K})^2 (1-\frac{2\beta \tilde{H}_{_{\rm D}}}{\kappa_{0}(1+\tilde{H}_{_{\rm D}}^2)^{3/2}}),
\label{rotating_velocity}
\end{equation}
where $\Omega_{\rm K}$ is Keplerian angular velocity of disk at radius $r$, $\kappa_{0}$ is the slope of magnetic field lines at the disk surface, and $\tilde{H}_{_{\rm D}}$ is defined as the ratio of disk half thickness to disk radius. We set $\tilde{H}_{_{\rm D}}=0.01$ here, so that ${\bf v}_{\phi}(r,\pi/2)$ is very close to the Keplerian rotating velocity.

In the simulation domain, the initial physical variables are set under the assumption of isothermal hydrostatic equilibrium in the vertical direction. The initial density distribution is given by
\begin{equation}
\rho(r,\theta)=\rho_{_{\rm C}} {\rm exp}(-\frac{GM_{_{\rm BH}}}{2c^2_{\rm s,_{C}}r(1-r_{\rm s}/r)^2{\rm tan}^2(\theta)}).
\label{density_distribution}
\end{equation}
Equation (\ref{density_distribution}) is slightly different from Yang et al. (2018) because we apply the pseudo-Newtonian potential (Paczy\'{n}sky \& Wiita 1980). The initial temperature $T(r,\theta)$ is set to be $T_{_{\rm C}}$. For the initial velocity, ${\bf v}_r(r,\theta)$ and ${\bf v}_{\theta}(r,\theta)$ are set to be null. The rotational velocity is given as ${\bf v}_\phi(r,\theta)=\sqrt{GM_{_{\rm BH}}/r}{\rm sin}(\theta)r/(r-r_{\rm s})$, which meets the equilibrium between the BH gravity and the centrifugal force.

The computational domain is discretized into $144\times160$ zones, whose size ratios are $(\bigtriangleup r)_{i+1} / (\bigtriangleup r)_{i} = 1.04$ in the radial direction and $(\bigtriangleup \theta)_{j+1} / (\bigtriangleup \theta)_{j} = 0.970072$ in the angular direction, respectively. The smallest radial size of zones is $\Delta r =0.0123r_{\rm s}$ at the inner boundary. The smallest angular size is $\Delta \theta =0^{o}.023589$ at $\theta = \pi/2$. The outflow boundary condition is employed at the inner and outer radial boundaries. At the pole (i.e. $\theta=0$), the axially symmetric boundary condition is applied. At the equator (i.e. $\theta=\pi/2$), density and temperature are set to be $\rho_{_{\rm C}}$ and $T_{_{\rm C}}$ at all times, the radial and vertical velocities are set to be null, and the rotating velocity is given by equation (\ref{rotating_velocity}).

\begin{figure*}
\scalebox{0.40}[0.50]{\rotatebox{0}{\includegraphics[bb=80 350 520 700]{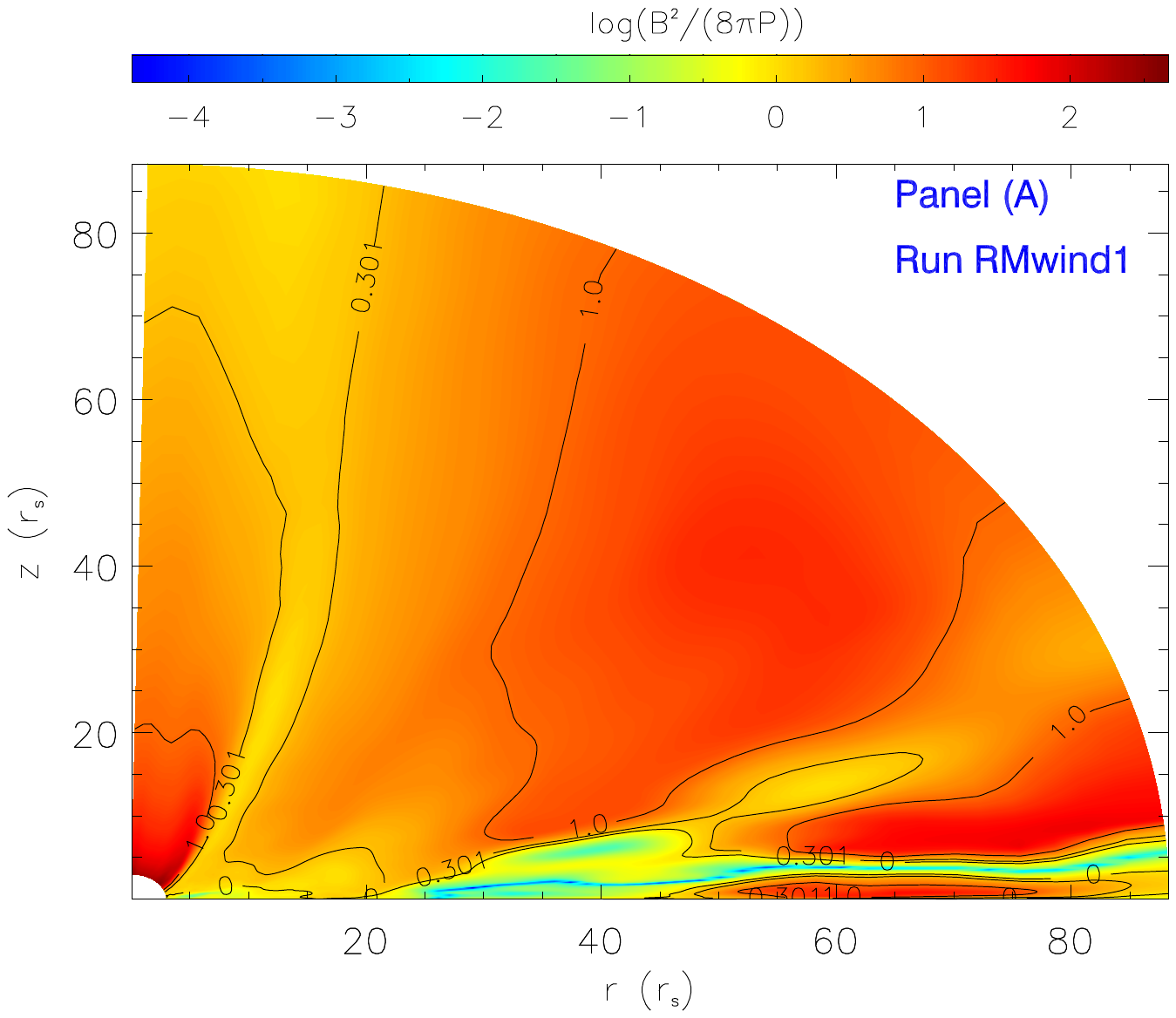}}}
\scalebox{0.40}[0.50]{\rotatebox{0}{\includegraphics[bb=90 350 520 700]{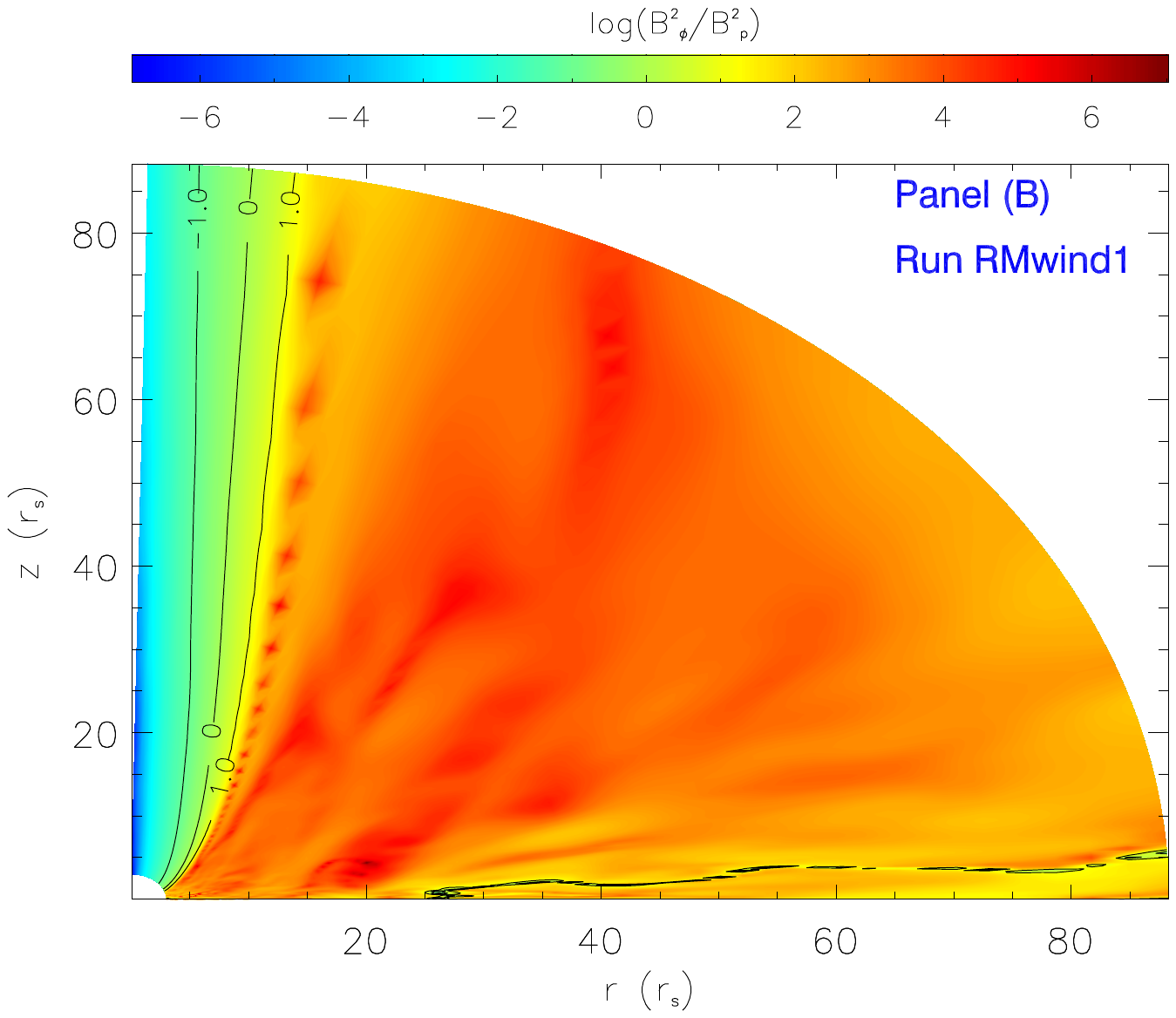}}}
\scalebox{0.40}[0.50]{\rotatebox{0}{\includegraphics[bb=90 350 520 700]{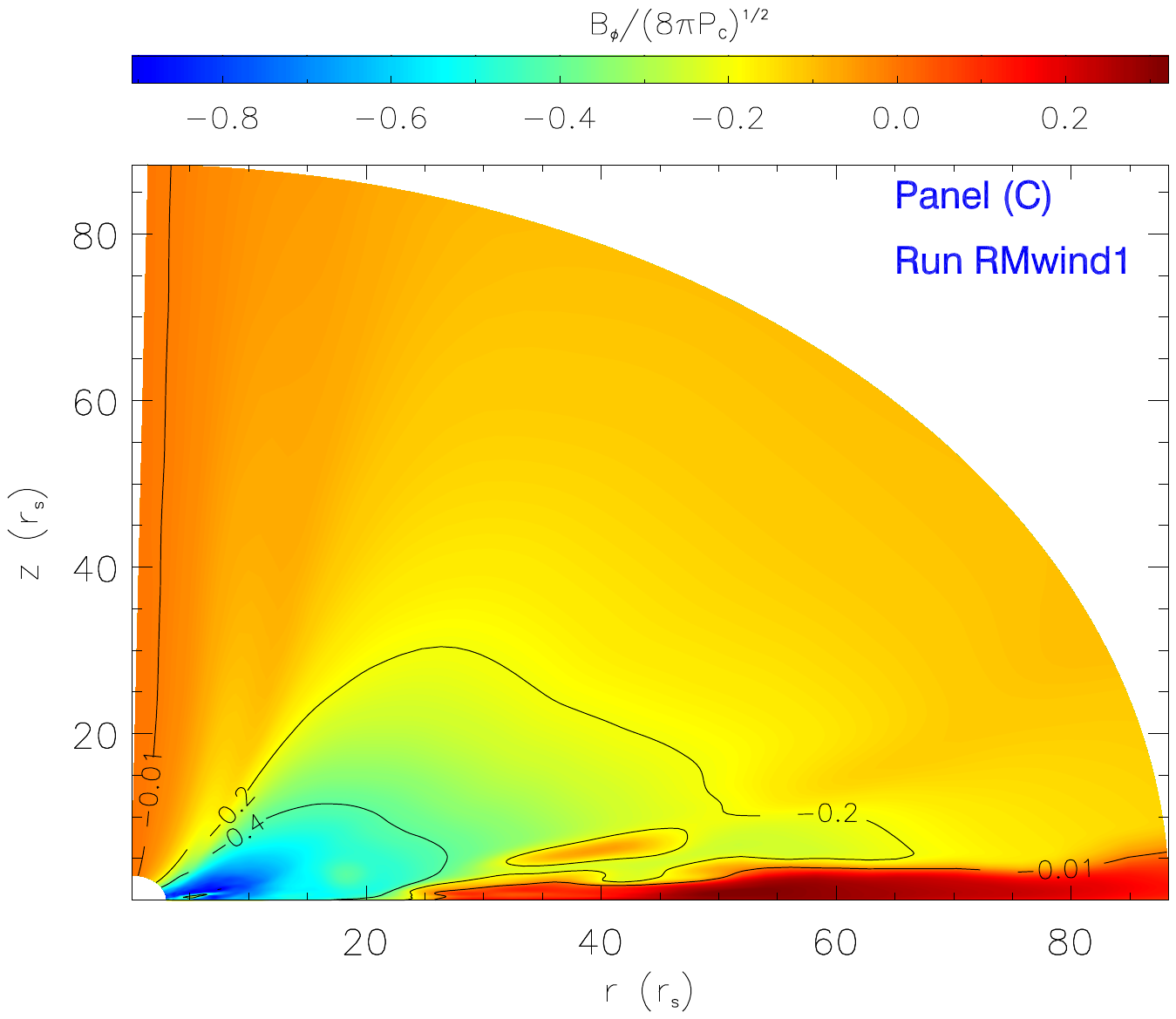}}}

\scalebox{0.40}[0.50]{\rotatebox{0}{\includegraphics[bb=80 340 520 700]{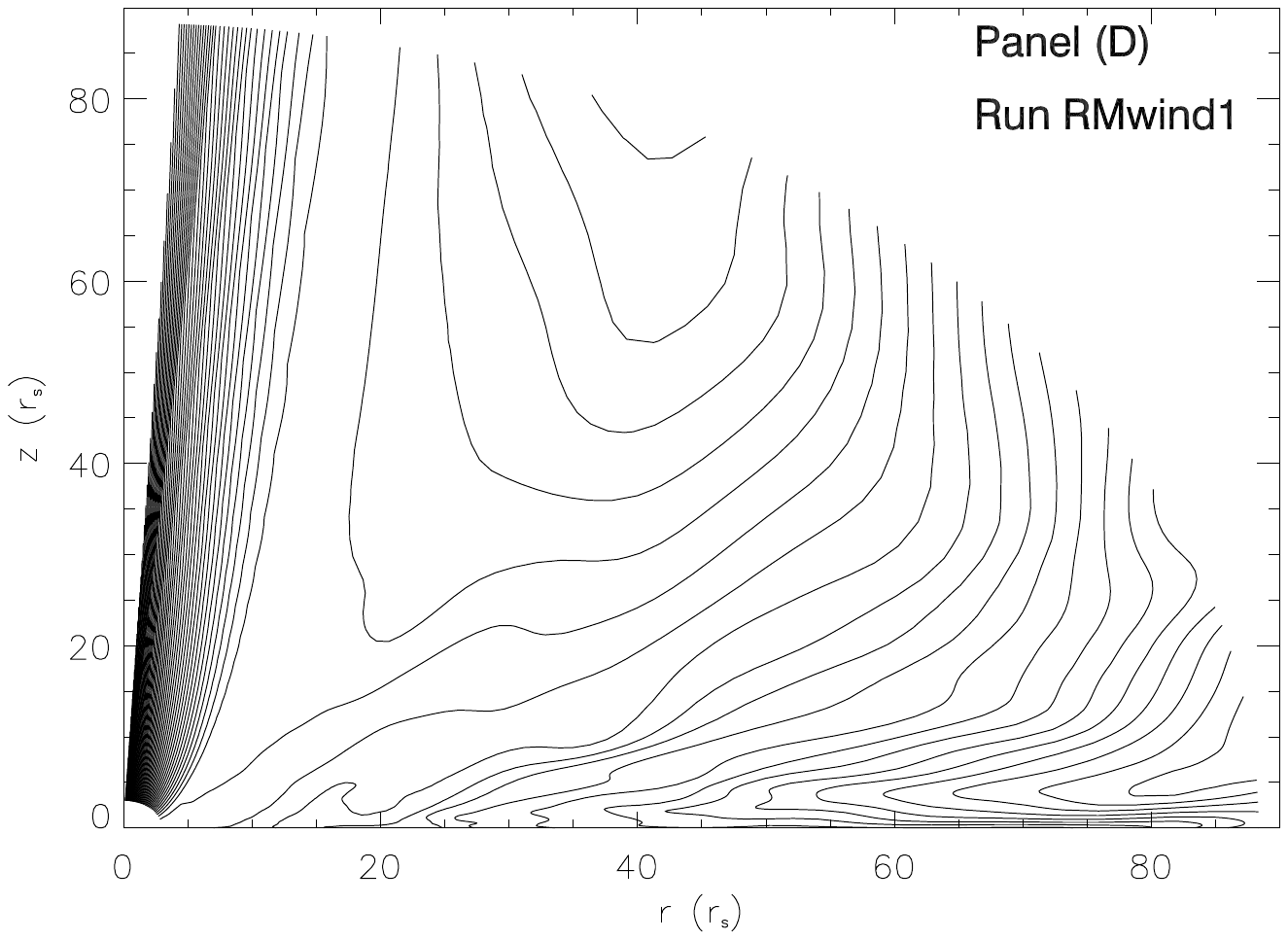}}}
\scalebox{0.40}[0.50]{\rotatebox{0}{\includegraphics[bb=90 340 520 700]{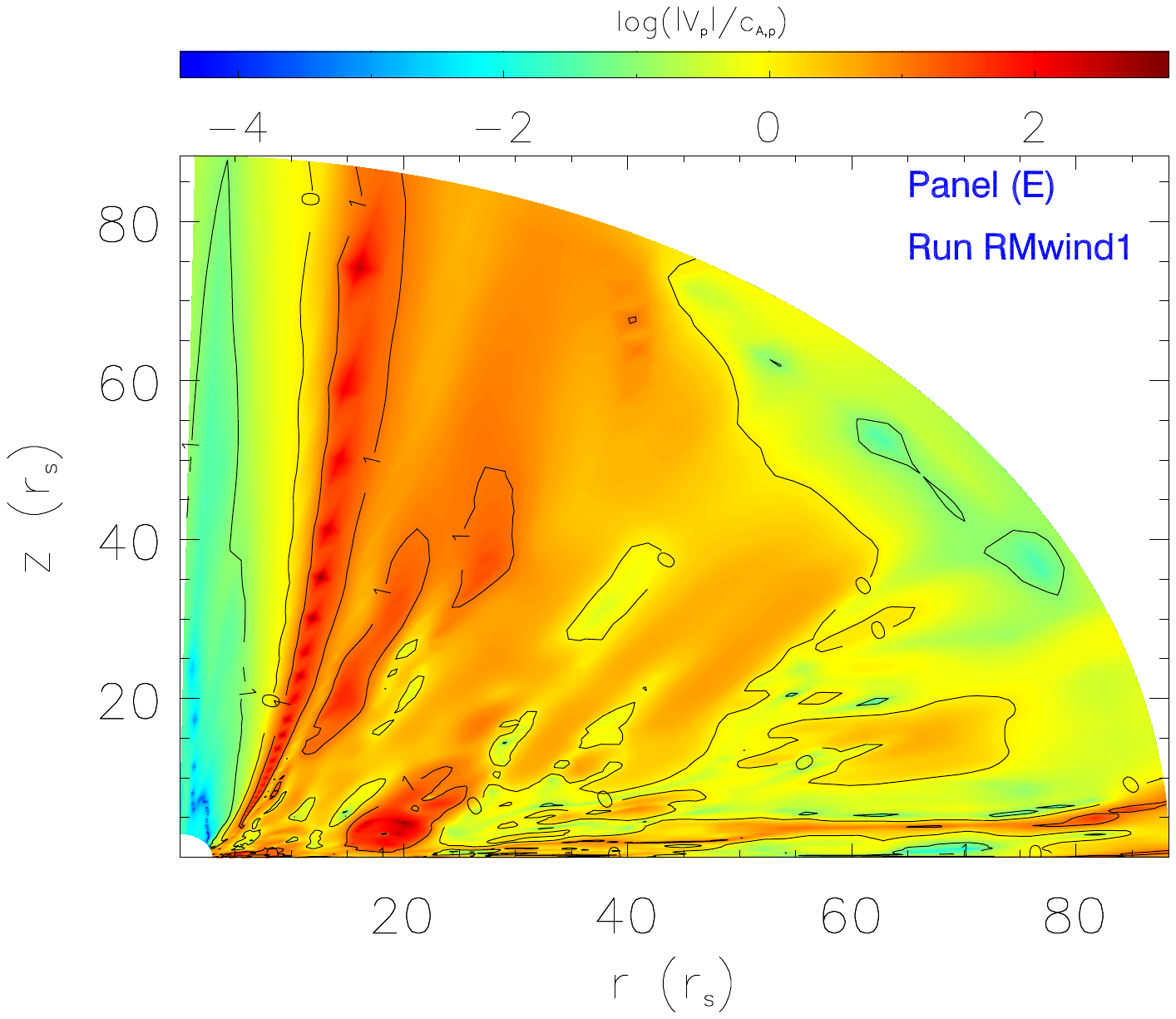}}}
\scalebox{0.40}[0.50]{\rotatebox{0}{\includegraphics[bb=90 340 520 700]{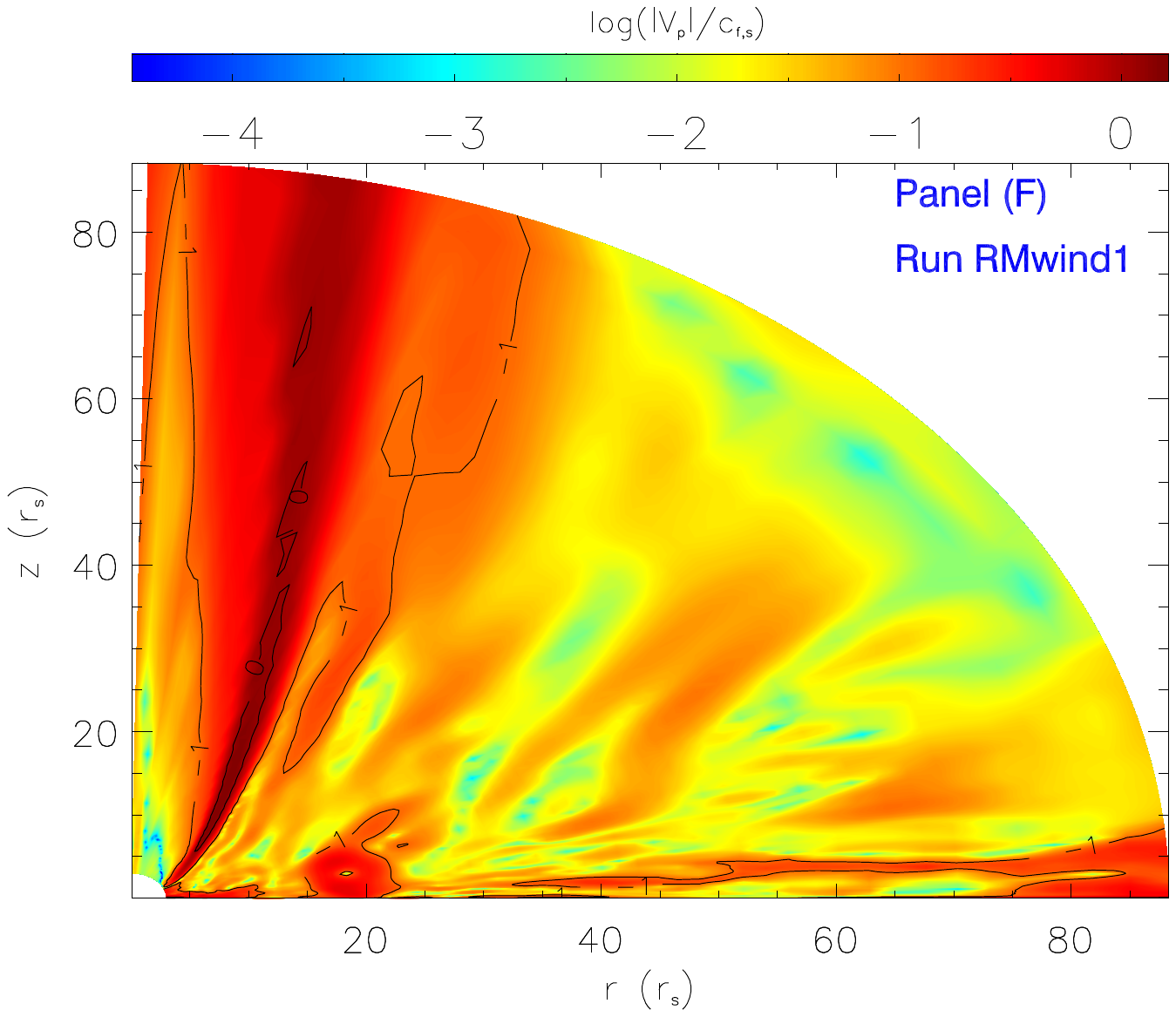}}}

\ \centering \caption{Series of time-averaged variables of run RMwind1 over $t=2.0$--3.0 $T_{\rm orb}$. Panel (A) shows the logarithm of the ratio ($\beta$) of magnetic pressure ($B^2/8\pi$) to gas pressure ($P$). The $log(\beta)$ contours are for 0, 0.301, and 1.0 (i.e. $\beta$=1,2 and 10, respectively). Panel (B) shows the logarithm of the ratio ($B^2_{\phi}/B^2_{\rm p}$) of toroidal ($B^2_{\phi}/8\pi$) to poloidal ($B^2_{\rm p}/8\pi$) magnetic pressure. The $log(B^2_{\phi}/B^2_{\rm p})$ contours are for -1, 0, and 1. Panel (C) shows the toroidal magnetic field in units of $\sqrt{8\pi P_{_{C}}}$, where $P_{_{\rm C}}$ is constant (i.e. the gas pressure of corona at the boundary of $\theta=\pi/2$ and $r=r_{\star}$). The $B_{\phi}/\sqrt{8\pi P_{_{\rm C}}}$ contours are for -0.4, -0.2, and -0.01. Panel (D) shows the magnetic field lines on the $r$--$z$ plane. Panels (E) and (F) show the ratio of the gas poloidal velocity ($|v_{\rm p}|$) to the poloidal component of the Alfv\'{e}n velocity and the ratio of the gas poloidal velocity to the fast magnetosonic velocity in the poloidal direction, respectively. The contours in panels (E) and (F) are also for -1, 0, and 1. }

\label{fig_4}
\end{figure*}

\begin{figure*}

\scalebox{0.40}[0.50]{\rotatebox{0}{\includegraphics[bb=80 350 520 700]{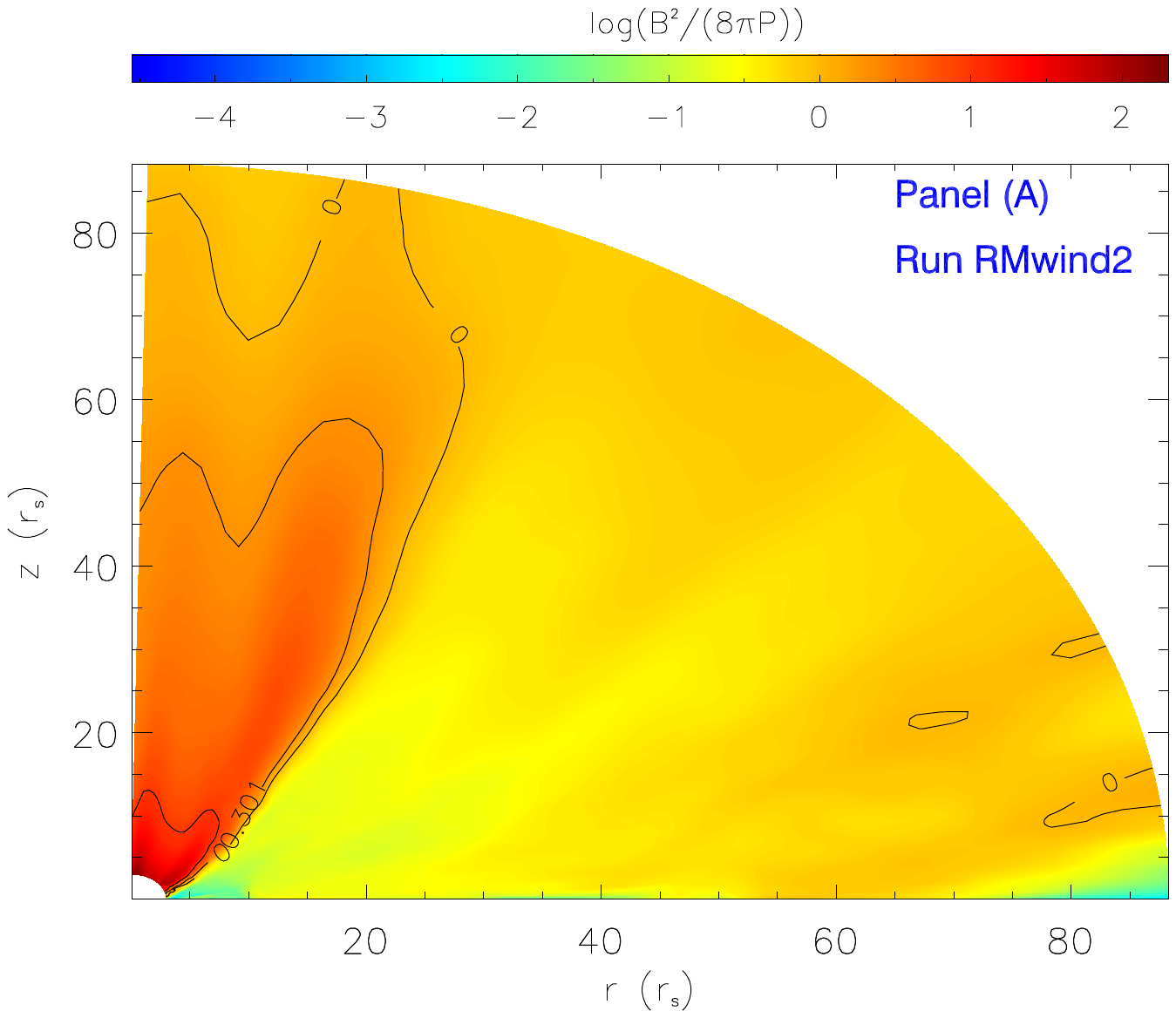}}}
\scalebox{0.40}[0.50]{\rotatebox{0}{\includegraphics[bb=90 350 520 700]{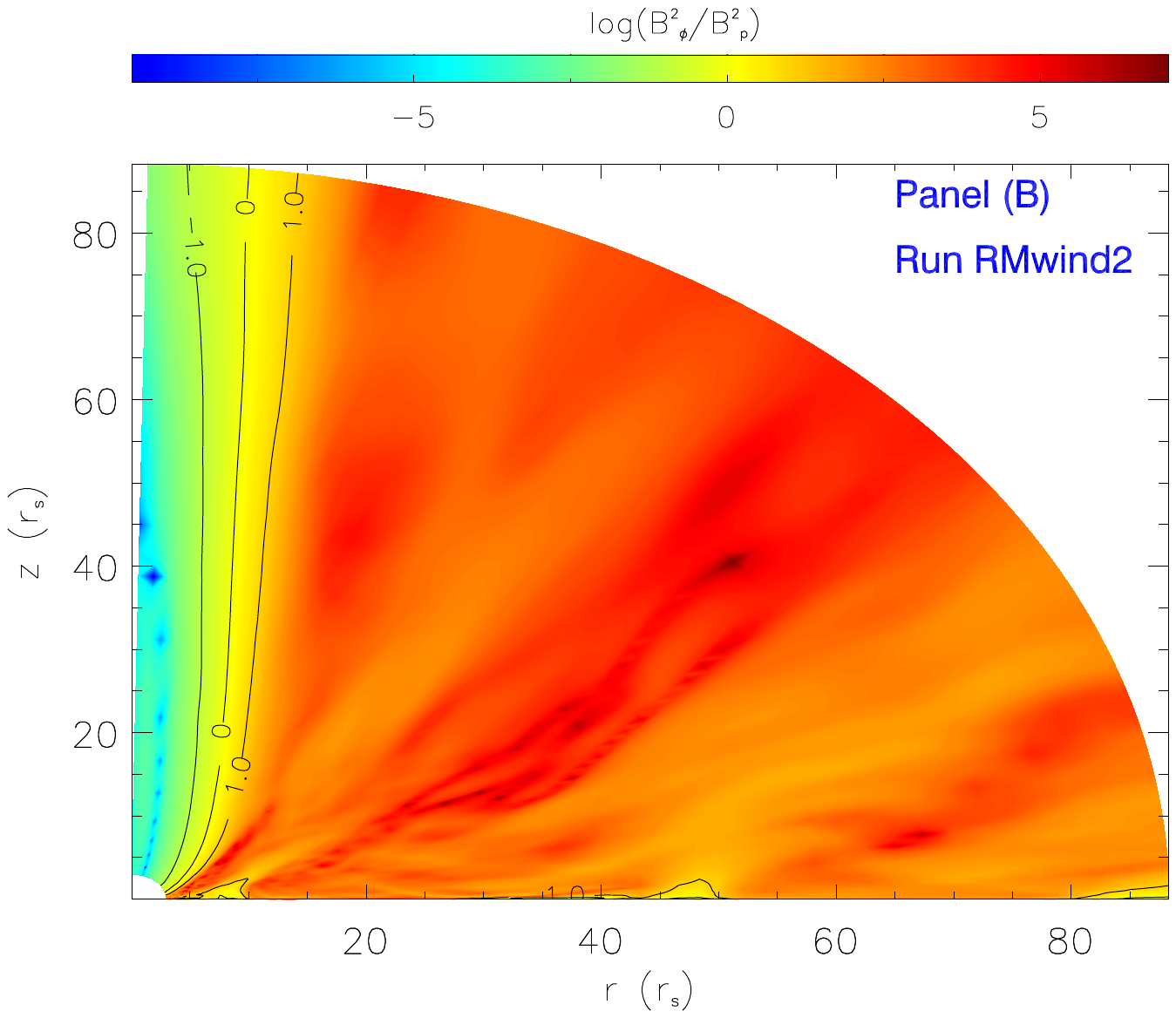}}}
\scalebox{0.40}[0.50]{\rotatebox{0}{\includegraphics[bb=90 350 520 700]{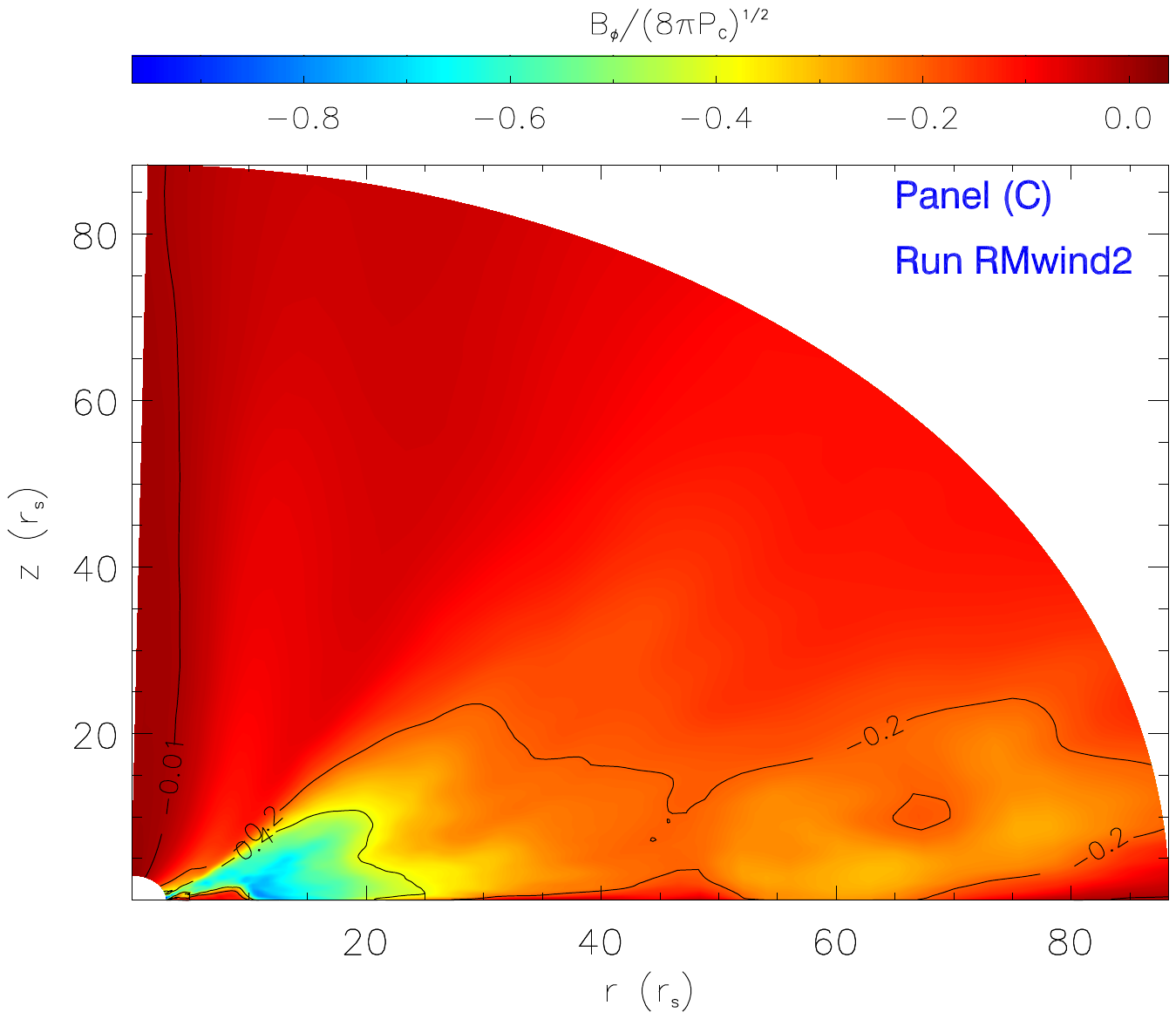}}}

\scalebox{0.40}[0.50]{\rotatebox{0}{\includegraphics[bb=80 340 520 700]{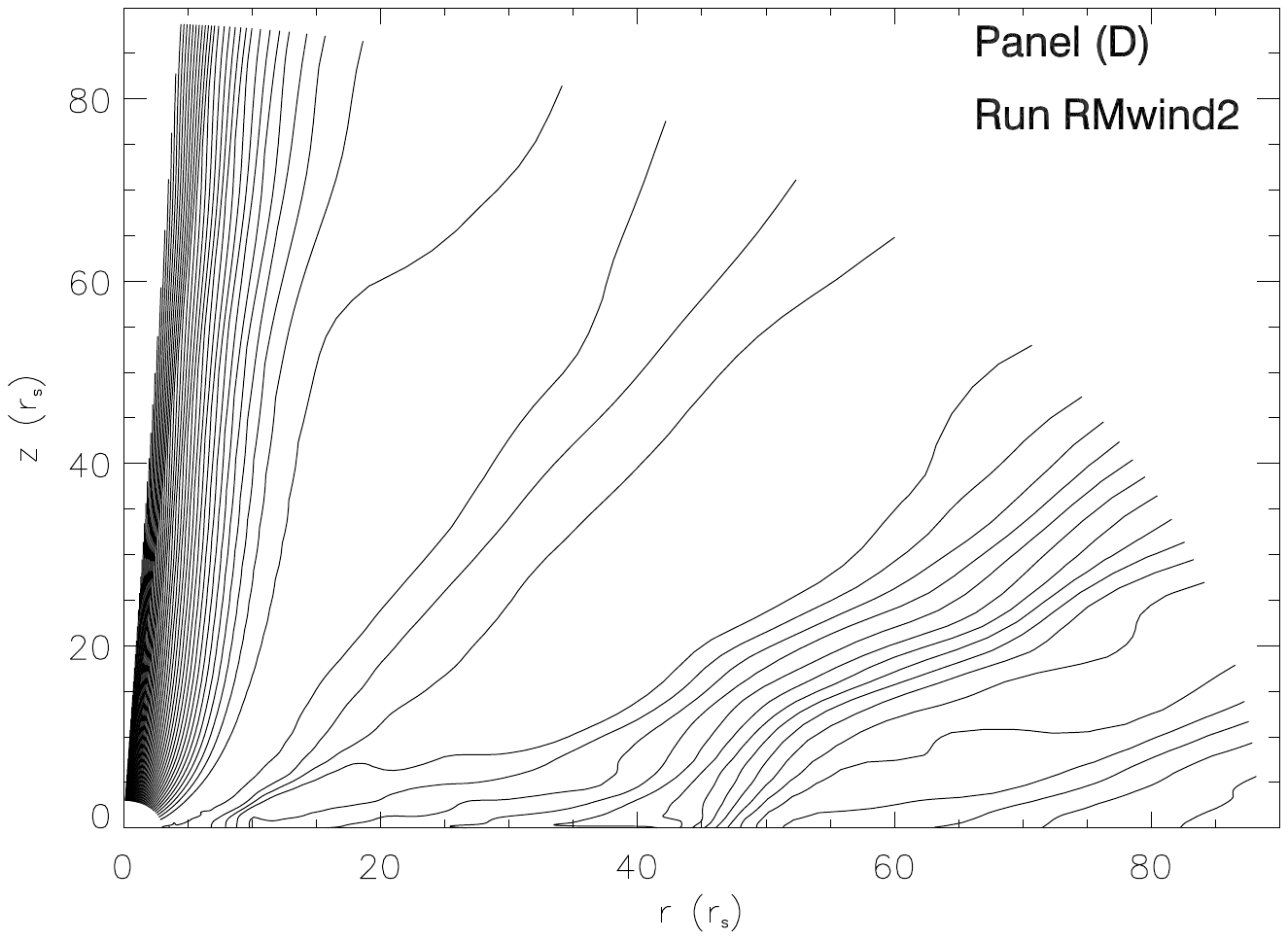}}}
\scalebox{0.40}[0.50]{\rotatebox{0}{\includegraphics[bb=90 340 520 700]{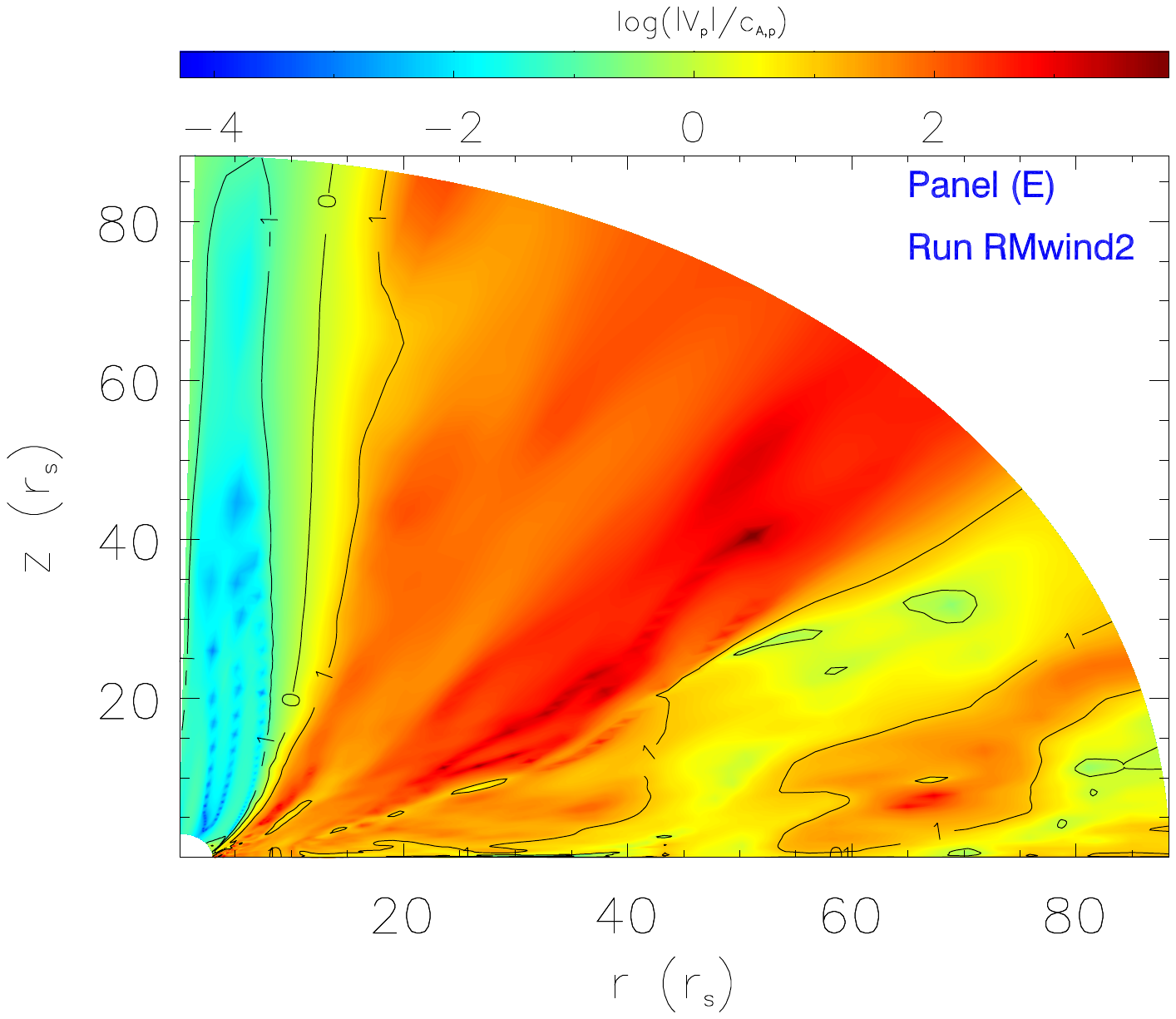}}}
\scalebox{0.40}[0.50]{\rotatebox{0}{\includegraphics[bb=90 340 520 700]{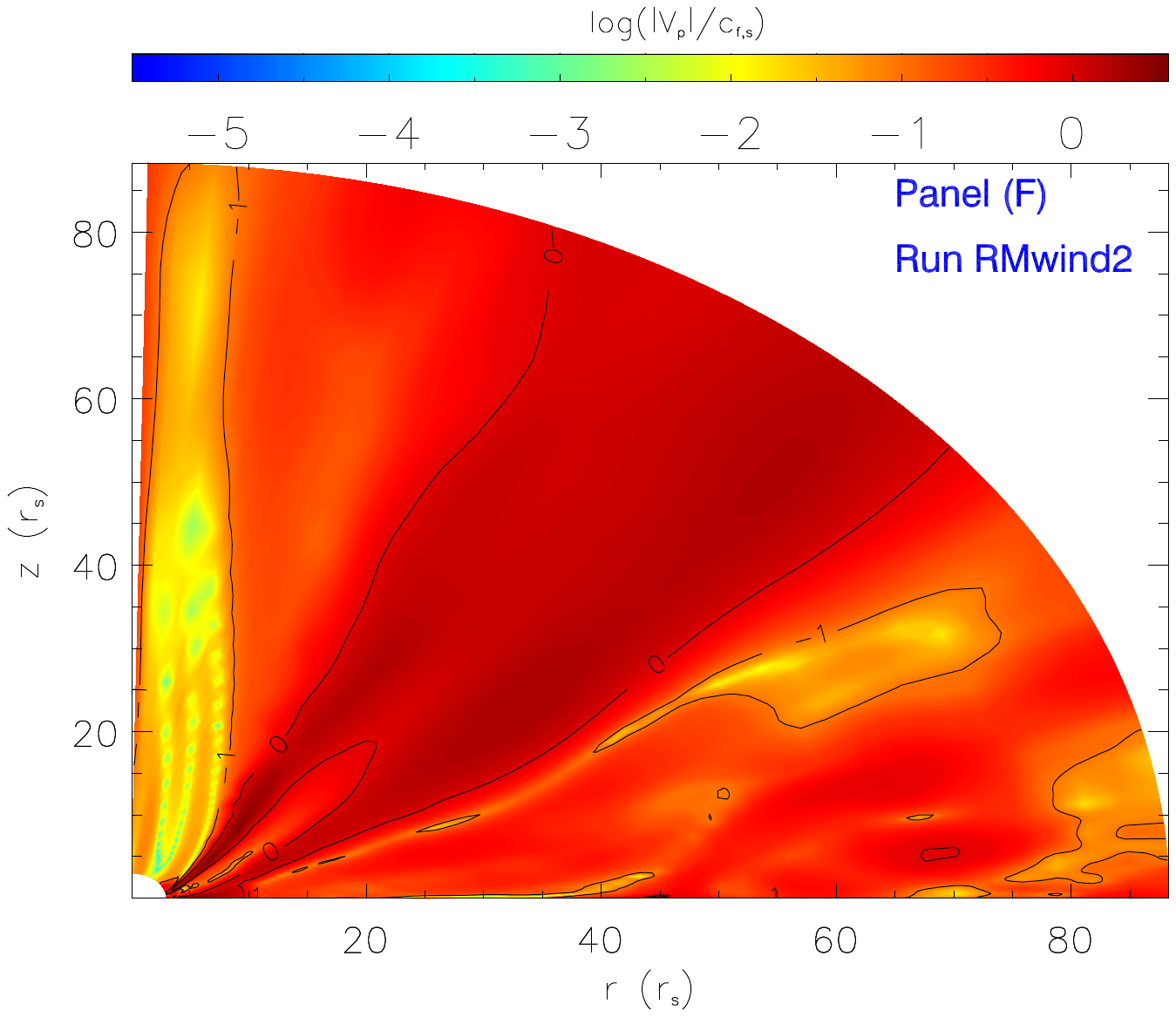}}}

\ \centering \caption{Series of time-averaged variables of run RMwind2 over $t=2.0$--$3.0$ $T_{\rm orb}$. Panels are the same as those in Figure \ref{fig_4}.}

 \label{fig_5}
\end{figure*}

In all of our simulations, we choose the units of length ($l_{0}$), velocity ($\rho_{0}$), and density ($\rho_{0}$) as follows:
\begin{equation}
l_{0}=\frac{6GM_{\rm bh}}{c^2}=8.86\times 10^{13}(\frac{M_{\rm bh}}{10^8 M_{\odot}}) \text{  cm},
\end{equation}
\begin{equation}
\upsilon_{0}=\sqrt{\frac{GM_{\rm bh}}{l_0}}=\frac{c}{\sqrt{6}}=1.22\times 10^{10} \text{  cm s}^{-1},
\end{equation}
\begin{equation}
\rho_{0}=n\times10^{-15}(\frac{M_{\rm bh}}{10^8 M_{\odot}})^{-1} \text{  g cm}^{-3},
 \label{rho_0}
\end{equation}
where $n$ is a free parameter. Under the optically thin assumption, equations (\ref{cont})--(\ref{induction_equation}) are density-free. In our simulations, the corona density ($\widetilde{\rho}_{_{\rm C}}$) at the equator is set to be $\widetilde{\rho}_{_{\rm C}}=1$ in units of $\rho_{0}$.

Except for the corona density ($\widetilde{\rho}_{_{\rm C}}$) and temperature ($T_{_{\rm C}}$), the model parameter space is made up of three parameters $\beta_0$, $\alpha_0$ and $\Gamma$. When $\beta_{0}=0$, the effect of magnetic field is ignored. When the effect of magnetic field is considered, $\beta_0$ is set to $10^{-2}$. This means that the initial ratio of magnetic pressure to gas pressure is $10^{-2}$ at the disk surface and at the inner boundary. We also set $\alpha_0=5.0$. This value means that the inclination angle (with respect to the disk surface) of initial magnetic field lines  is $50^o.5$ at the disk surface and at the inner boundary. In addition, we set the density floor to be $10^{-6}$ in units of $\rho_{0}$ and the mean molecular weight ($\mu$) to be 0.5.

\section{Results}
Our models are summarized in Table 1. Runs Rwind1 and Rwind2 in Table 1 do not include the effect of magnetic field, while the other runs include the effect of magnetic field. All of the models reach a quasi-steady state after $\sim0.5$ $T_{\rm orb}$. In this paper, we stop the simulations at 3 $T_{\rm orb}$. Figure \ref{fig_1} shows the time evolution of mass outflow rate at the outer boundary for runs RMwind1 and RMwind2. It is clear that after $\sim0.5$ $T_{\rm orb}$, the mass outflow rate in each model fluctuates around a mean value. A quasi-steady state has been achieved in these two models.

Table 1 also gives the mass outflow ($\dot{M}_{\rm out}$) as well as the kinetic ($P_{\rm k,out}$) and thermal energy ($P_{\rm th,out}$) flux carried out by the outflows at the outer boundary. They are time-averaged on the time range of 2.0-3.0 $T_{\rm orb}$, where $T_{\rm orb}$ is the Keplerian orbital period at 90 $r_{\rm s}$. In the model with low luminosity ($\Gamma=0.25$), magnetic field is helpful to increase the mass outflow rate and significantly increase $P_{\rm k,out}$ and $P_{\rm th,out}$. In the model with high luminosity ($\Gamma=0.75$), after considering magnetic field, $\dot{M}_{\rm out}$, $P_{\rm k,out}$, and $P_{\rm th,out}$ just slightly increase. We will give detailed explanations below.

\subsection{Magnetized Outflows (or Winds)}
Figures \ref{fig_2} and \ref{fig_3} show the time-averaged density, Mach number, and poloidal velocity (arrows), respectively. For the sake of comparison, the models (runs Rwind1 and Rwind2) without magnetic field are also plotted in Figures \ref{fig_2} and \ref{fig_3}. The outflows driven only by radiation force have been studied in Yang et al. (2018). In this paper, the boundary condition at the equator is different from that used in Yang et al (2018), so that the outflow properties of runs Rwind1 and Rwind2 are different from those of outflows found in Yang et al (2018). However, we mainly study the magnetized outflows in this paper. The time-averaged values in Figures \ref{fig_2} and \ref{fig_3} are obtained by averaging 100 output files over $t=2.0$--$3.0$ $T_{\rm orb}$. In panel (A), the length of arrows is proportional to $|v_{\rm p}| (\equiv\sqrt{v_{r}^2+v_{\theta}^2})$. In panel (B), the blue solid line overplotted marks the surface of $|v_{\rm p}|=c_{\rm s}$. $c_{\rm s}\equiv\sqrt{\gamma P/\rho}$ is the sonic speed of gas.

Initially, the gas is in force balance, where the centrifugal force balances the horizontal component of BH gravity and the pressure gradient force balances the vertical component of BH gravity. When an extra force, such as the radiation force from a disk, is exerted on the gas, outflows are easily launched. However, if the radiation force due to Compton scattering is too weak, the ordered outflows cannot be effectively driven the from the corona (Yang et al. 2018). For run RMwind1 with low luminosity ($\Gamma=0.25$), Figure \ref{fig_2} shows that an ordered outflow exists in the range of $\theta=\sim 0^{o}$--$30^{o}$, while the gas in the range of $\theta=\sim 30^{o}$--$90^{o}$ is turbulent. For run Rwind1, the small-scale turbulence takes place around the equator and the ordered outflow is not observed. In run Rwind1, magnetic field is neglected, the radiation force is not strong enough to drive the middle- and high-latitude ordered outflows. In run RMwind1, magnetic field plays an important role in driving the ordered outflows. For run RMwind2 with high luminosity ($\Gamma=0.75$), Figure \ref{fig_3} shows that the ordered outflows are driven in the region of $\theta=\sim 0^{o}$--$60^{o}$, while turbulence exists in the region of $\theta=\sim 60^{o}$--$90^{o}$. For run Rwind2, we do not observe the outflow around the rotational axis. Comparing runs Rwind2 and RMwind2, we find that in these two models, the two-dimensional structures of the flow are similar except in the region around the rotational axis. Because the gas density is very low around the rotational axis, the contributions of the outflowing gas around the rotational axis to $\dot{M}_{\rm out}$, $P_{\rm k,out}$ and $P_{\rm th,out}$ are negligible. As shown in Figures \ref{fig_2} and \ref{fig_3}, in the two models with magnetic field, both a collimated outflow around rotational axis and a wide-angle ordered outflow away from rotational axis are observed. The collimated outflow has much higher outflowing radial velocity than that of the wide-angle ordered outflow. The wide-angle ordered outflows are supersonic in some regions. When the disk luminosity is high ($\Gamma=0.75$), gas can be blown away around the equator from the region within 10 $r_{s}$, where a hot corona of $10^9$ K exists. When the disk luminosity is low ($\Gamma=0.25$), turbulence motions dominate around the equator.

\begin{figure*}

\scalebox{0.35}[0.35]{\rotatebox{0}{\includegraphics[bb=45 20 500 360]{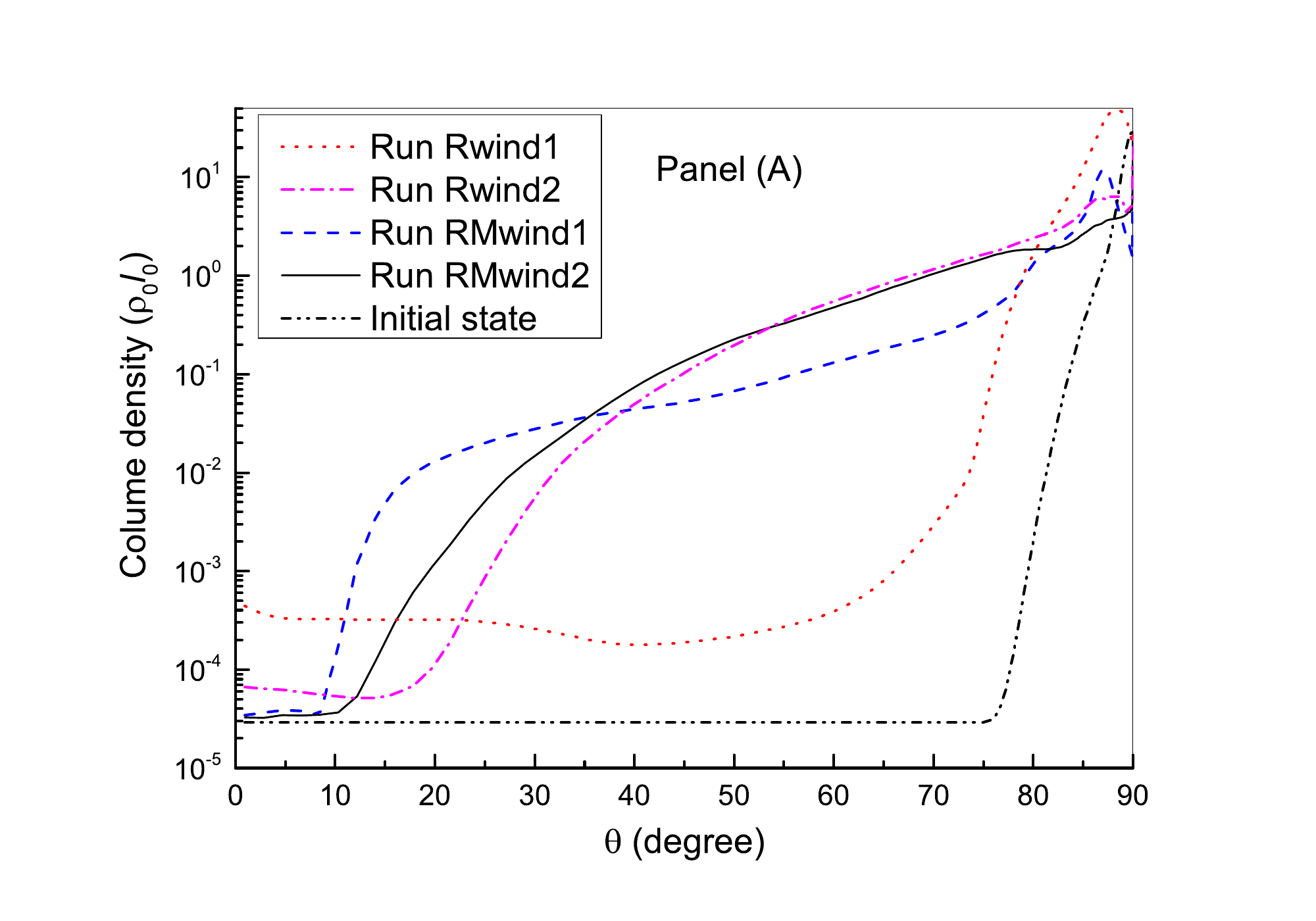}}}
\scalebox{0.35}[0.35]{\rotatebox{0}{\includegraphics[bb=45 20 500 360]{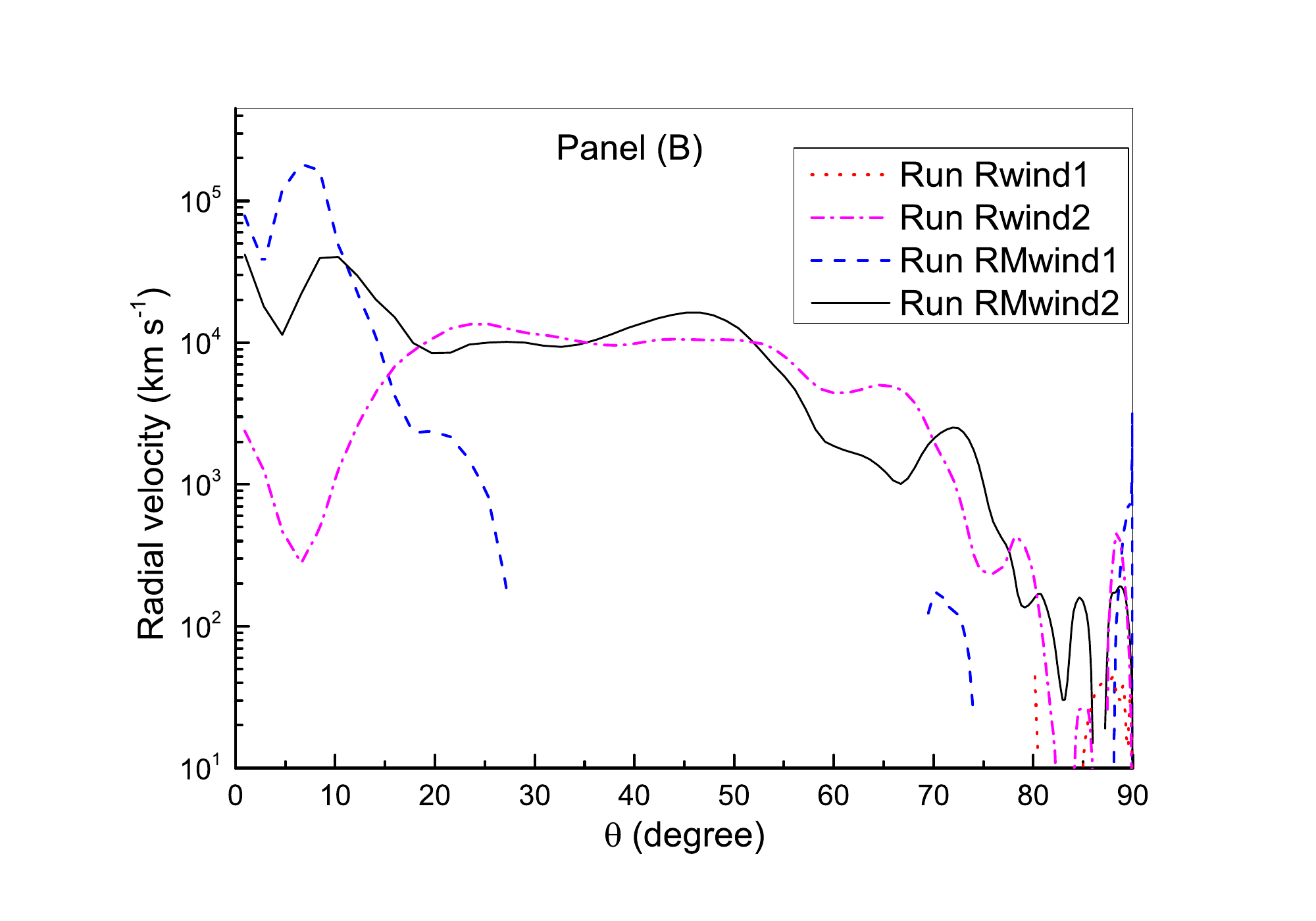}}}
\scalebox{0.35}[0.35]{\rotatebox{0}{\includegraphics[bb=45 20 500 360]{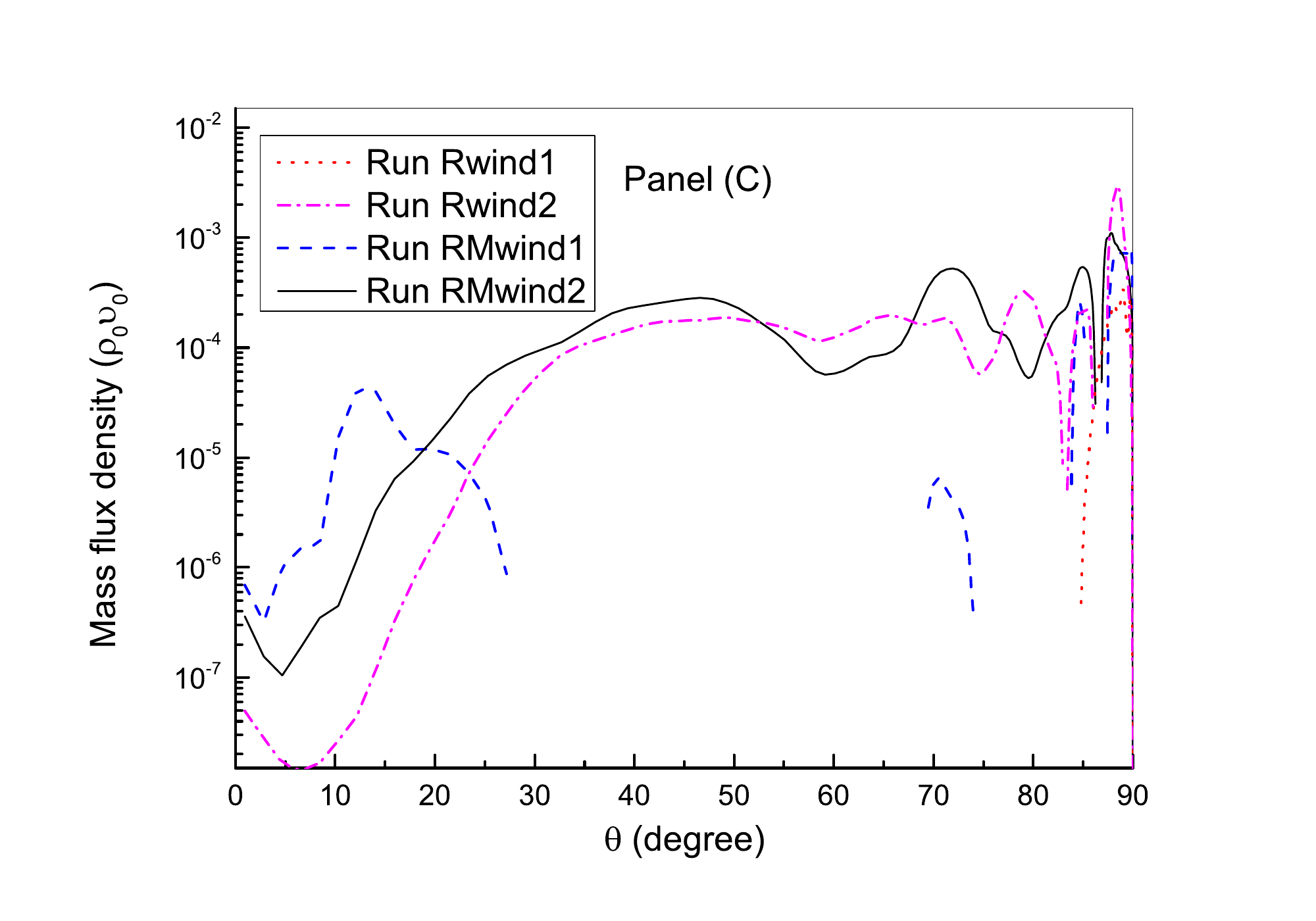}}}

\ \centering \caption{Angular profiles of a variety of time-averaged variables. Panel (A): column density; panel (B): radial velocity at the outer boundary; panel (C): mass flux density at the outer boundary.}
\label{fig_6}
\end{figure*}

\begin{figure*}

\scalebox{0.35}[0.35]{\rotatebox{0}{\includegraphics[bb=45 20 500 360]{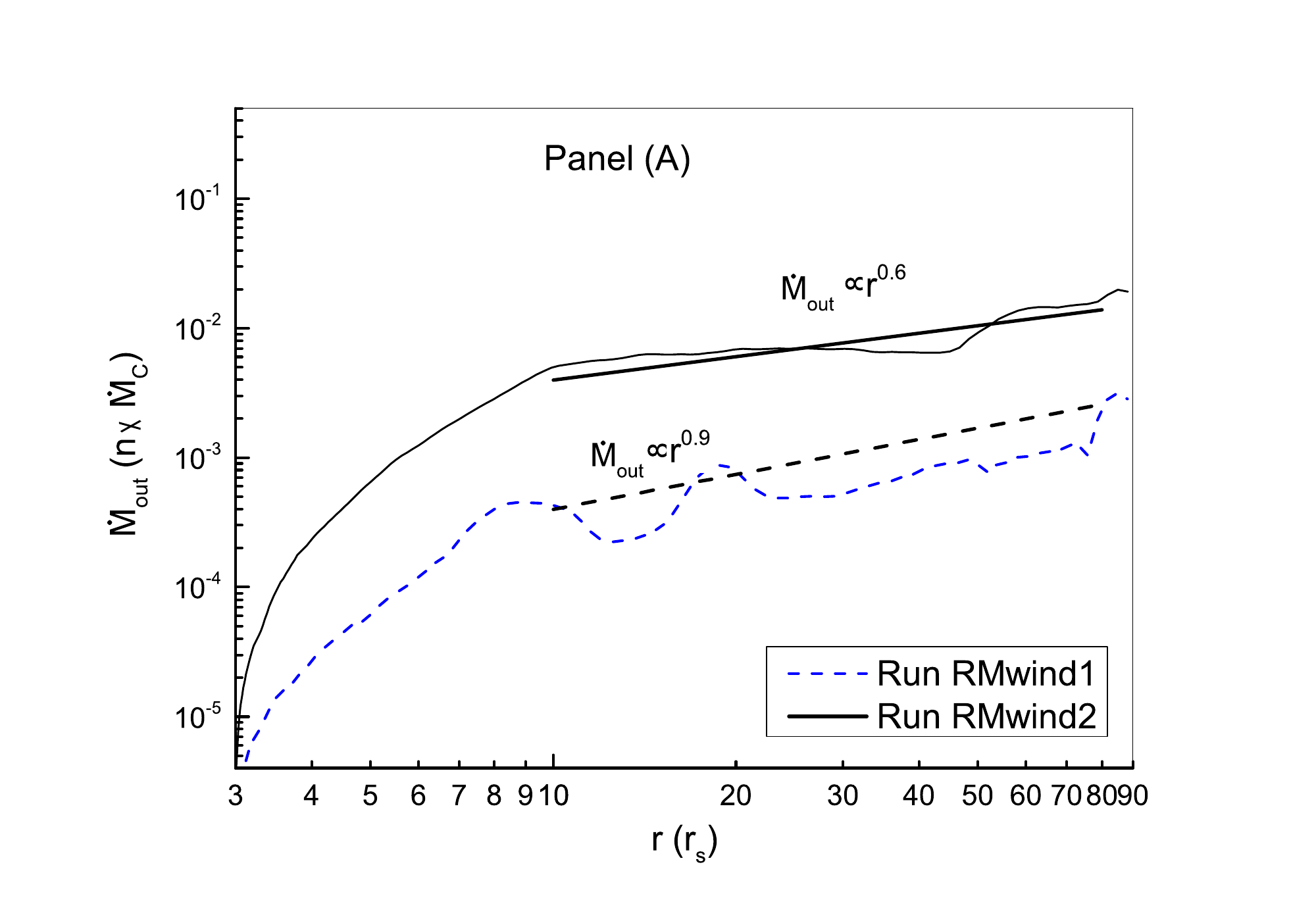}}}
\scalebox{0.35}[0.35]{\rotatebox{0}{\includegraphics[bb=45 20 500 360]{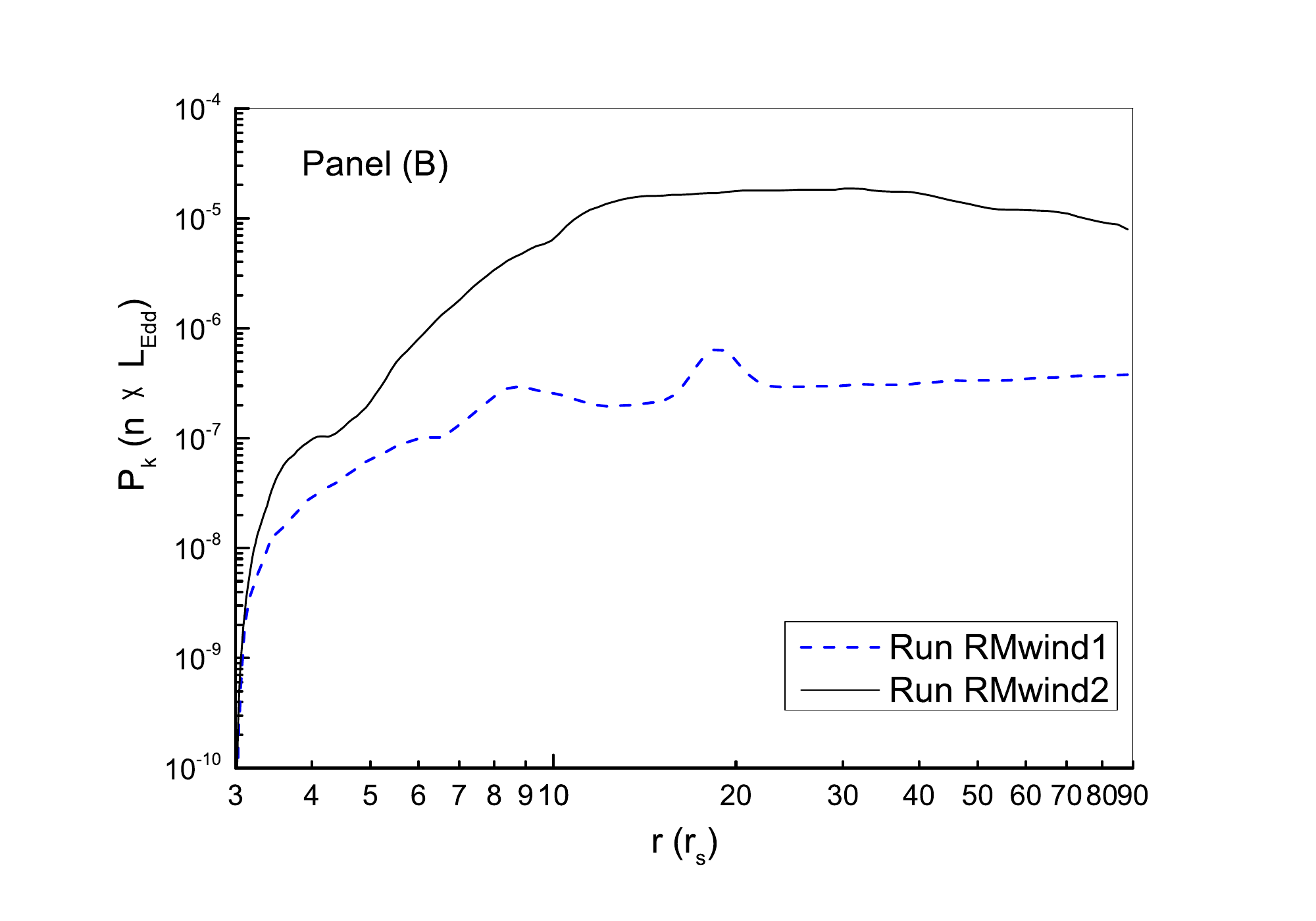}}}
\scalebox{0.35}[0.35]{\rotatebox{0}{\includegraphics[bb=45 20 500 360]{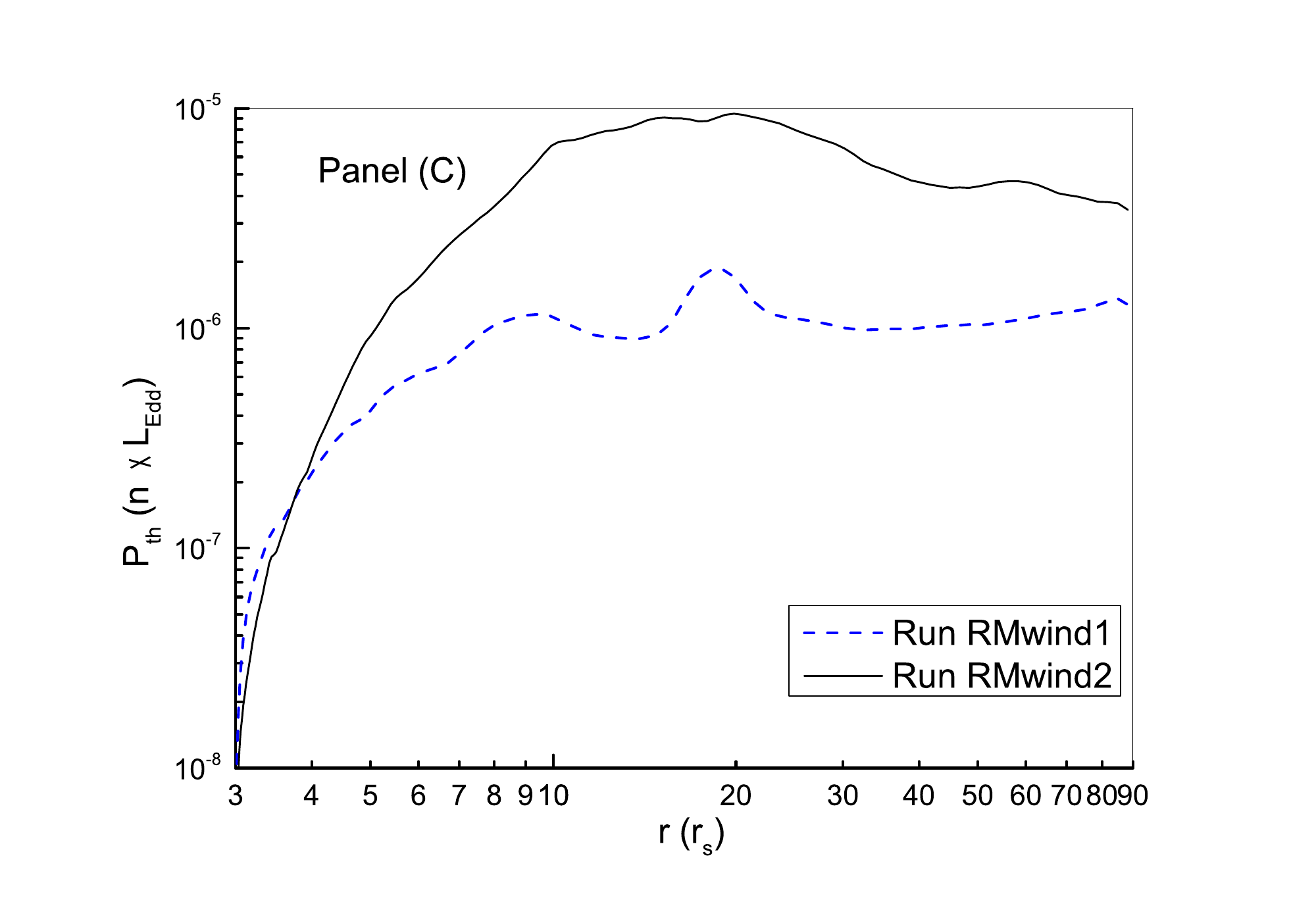}}}

\ \centering \caption{Time-averaged quantities as functions of $r$. Panel (A): outflow mass rate; panel (B): kinetic power of outflows; panel (C): thermal energy flux carried by outflows. In panel (A), the thick solid line and the thick dashed line are used to scale the radial dependence of outflow mass rate. The letter ``n'' is defined in Equation (12).}
\label{fig_7}
\end{figure*}

Initially, the magnetic field only has a poloidal component. Because of electromagnetic induction, magnetic field quickly evolves with time. During the evolution of magnetic field, $B_{\phi}$ is generated by rotation and the magnetic field lines on the $r$--$z$ plane are distorted. Figures \ref{fig_4} and \ref{fig_5} present two-dimensional structures of time-averaged variables, for runs RMwind1 and RMwind2. In Figures \ref{fig_4} and \ref{fig_5}, panel (A) shows the ratio ($\beta$) of magnetic pressure to gas pressure. Panel (B) gives the ratio ($B^2_{\phi}/B^2_{\rm p}$) of toroidal ($B^2_{\phi}/8\pi$) to poloidal ($B^2_{\rm p}/8\pi\equiv (B_{r}^2+B_{\theta}^2)/8\pi$) magnetic pressure. Panels (C) and (D) present the toroidal magnetic field and the magnetic field lines on the $r$--$z$ plane. Panels (E) and (F) show the ratio of the gas poloidal velocity ($|v_{\rm p}|\equiv\sqrt{v_{r}^2+v_{\theta}^2}$) to the poloidal component ($c_{\rm A,p}\equiv\sqrt{(B_{r}^2+B_{\theta}^2)/4\pi\rho}$) of the Alfv\'{e}n velocity ($c_{\rm A}\equiv\sqrt{(B_{r}^2+B_{\theta}^2+B_{\phi}^2)/4\pi\rho}$) and the ratio of the gas poloidal velocity to the fast magnetosonic velocity ($c_{\rm f,p}^2\equiv0.5[(c_{\rm s}^2+c_{\rm A}^2)+\sqrt{(c_{\rm s}^2+c_{\rm A}^2)^2-4c_{\rm s}^2c_{\rm A,p}}]$) in the poloidal direction, respectively.

In run RMwind1, panel (A) of Figure \ref{fig_4} shows that the magnetic pressure dominates over the gas pressure in almost all regions of the computational domain. For run RMwind2, panel (A) of Figure \ref{fig_5} shows that the magnetic pressure is stronger than the gas pressure in the region of $\theta<\sim23^{0}$. Magnetic pressure is weaker than the gas pressure in the region of $\theta>\sim23^{0}$. In the two models, the poloidal magnetic field is weaker than the toroidal magnetic field in most of the region of the computational domain, except in the region around the axis, as shown in panels (B) of Figures \ref{fig_4} and \ref{fig_5}. In the region around the axis, the poloidal magnetic field dominates over the toroidal magnetic field.

In the models with magnetic field, in most of the region of the computational domain, magnetic pressure is dominated by $B_{\phi}^2/8\pi$. In the $B_{\phi}^2$-dominated region, the gradient force of magnetic pressure points toward the direction of $B_{\phi}^2$ decreasing and is perpendicular to the $B_{\phi}^2$ contours shown in panel (C) of Figures \ref{fig_4} and \ref{fig_5}. Panel (E) of Figures \ref{fig_4} and \ref{fig_5} shows that the collimated outflow around the rotational axis moves with a velocity lower than the poloidal Alfv\'{e}n velocity, while the wide-angle outflow moves with a velocity larger than the poloidal Alfv\'{e}n velocity. For run RMwind1, the poloidal velocity of outflows is lower than the fast magnetosonic speed in most regions of the computational domain, as shown in panels (F) of Figures \ref{fig_4}. For run RMwind2, the outflows around $\theta=45^o$ move outwards with a velocity higher than the fast magnetosonic speed.

\begin{figure*}

\scalebox{0.28}[0.28]{\rotatebox{0}{\includegraphics[bb=58 33 500 360]{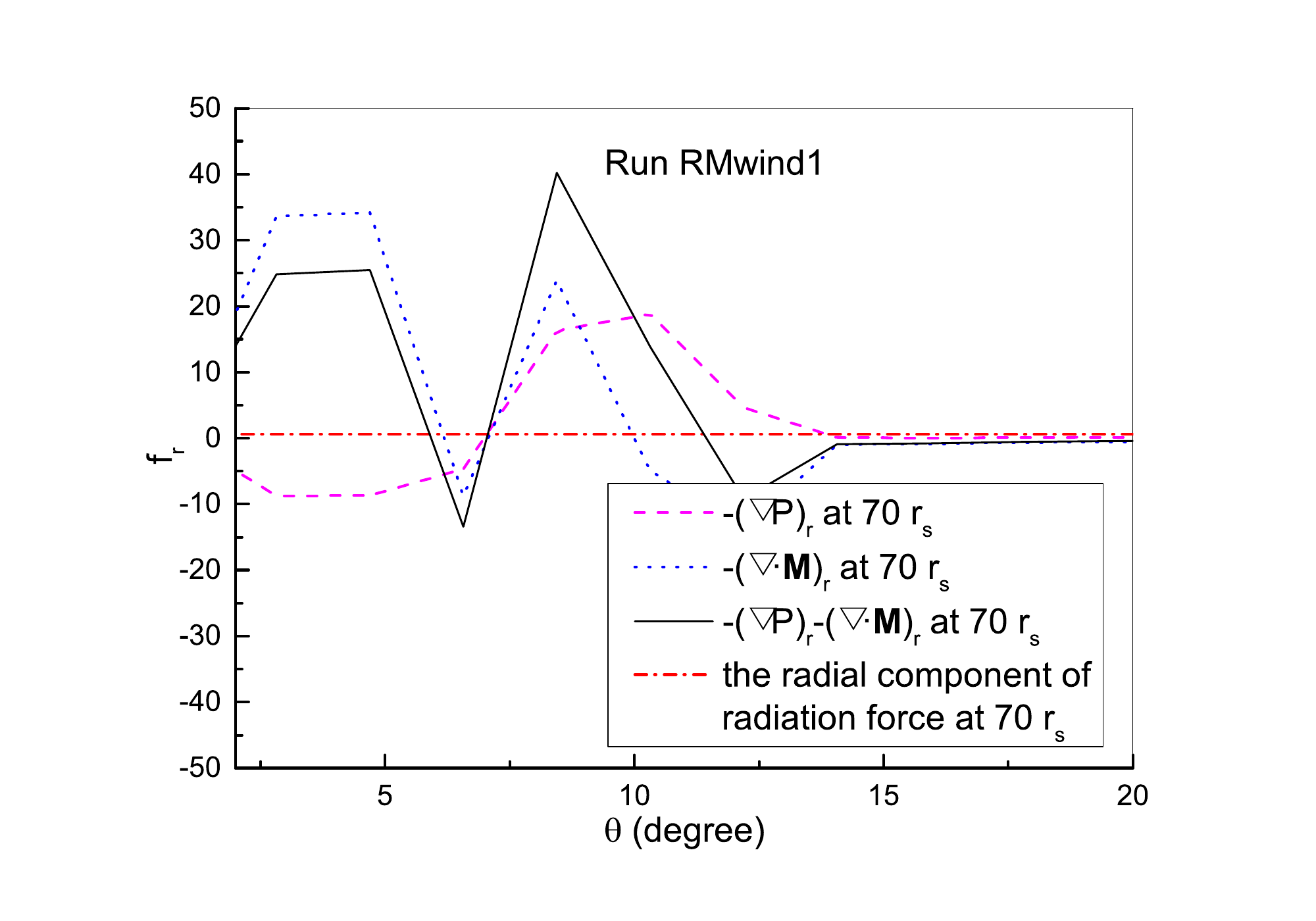}}}
\scalebox{0.28}[0.28]{\rotatebox{0}{\includegraphics[bb=58 33 500 360]{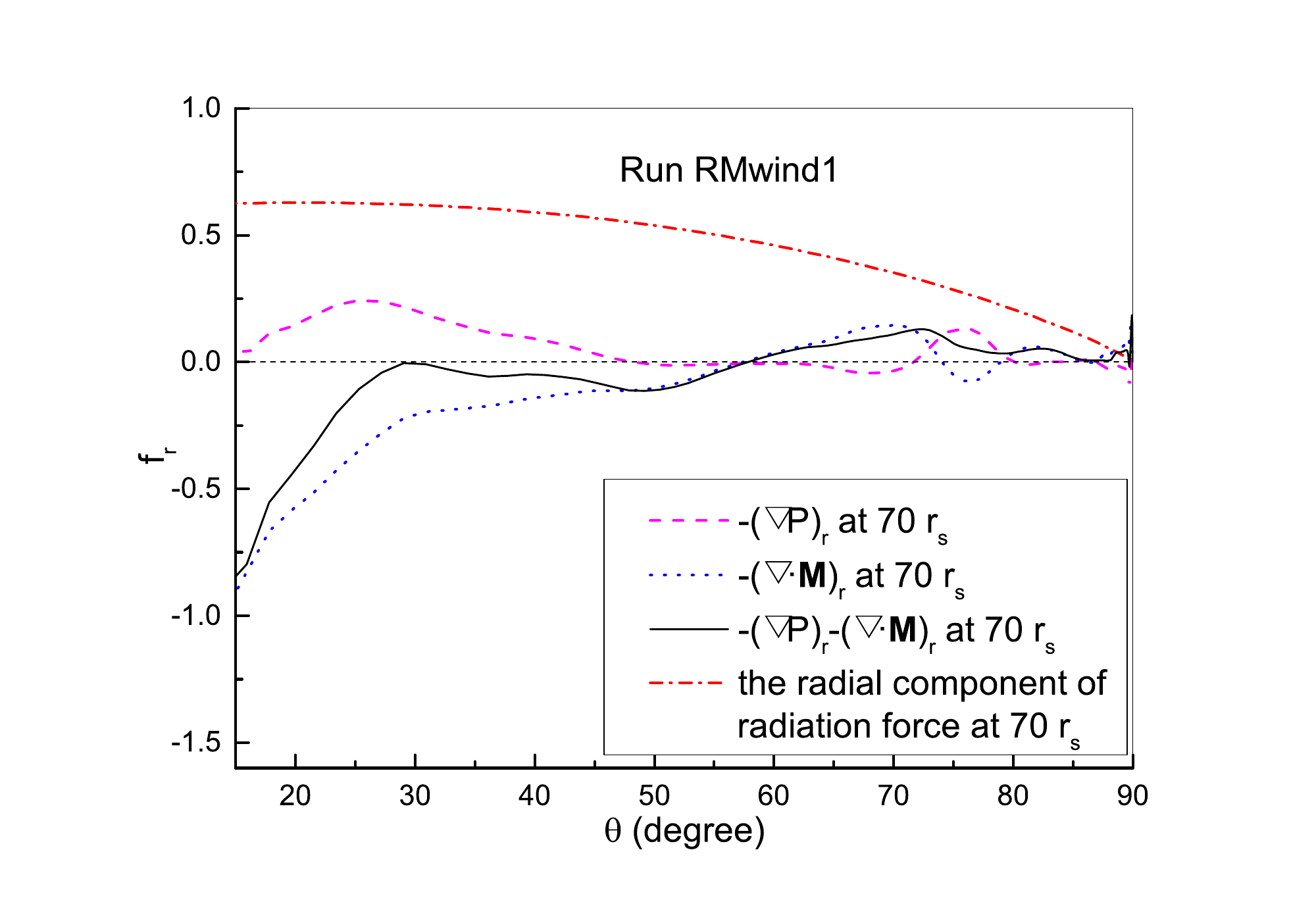}}}
\scalebox{0.28}[0.28]{\rotatebox{0}{\includegraphics[bb=58 33 500 360]{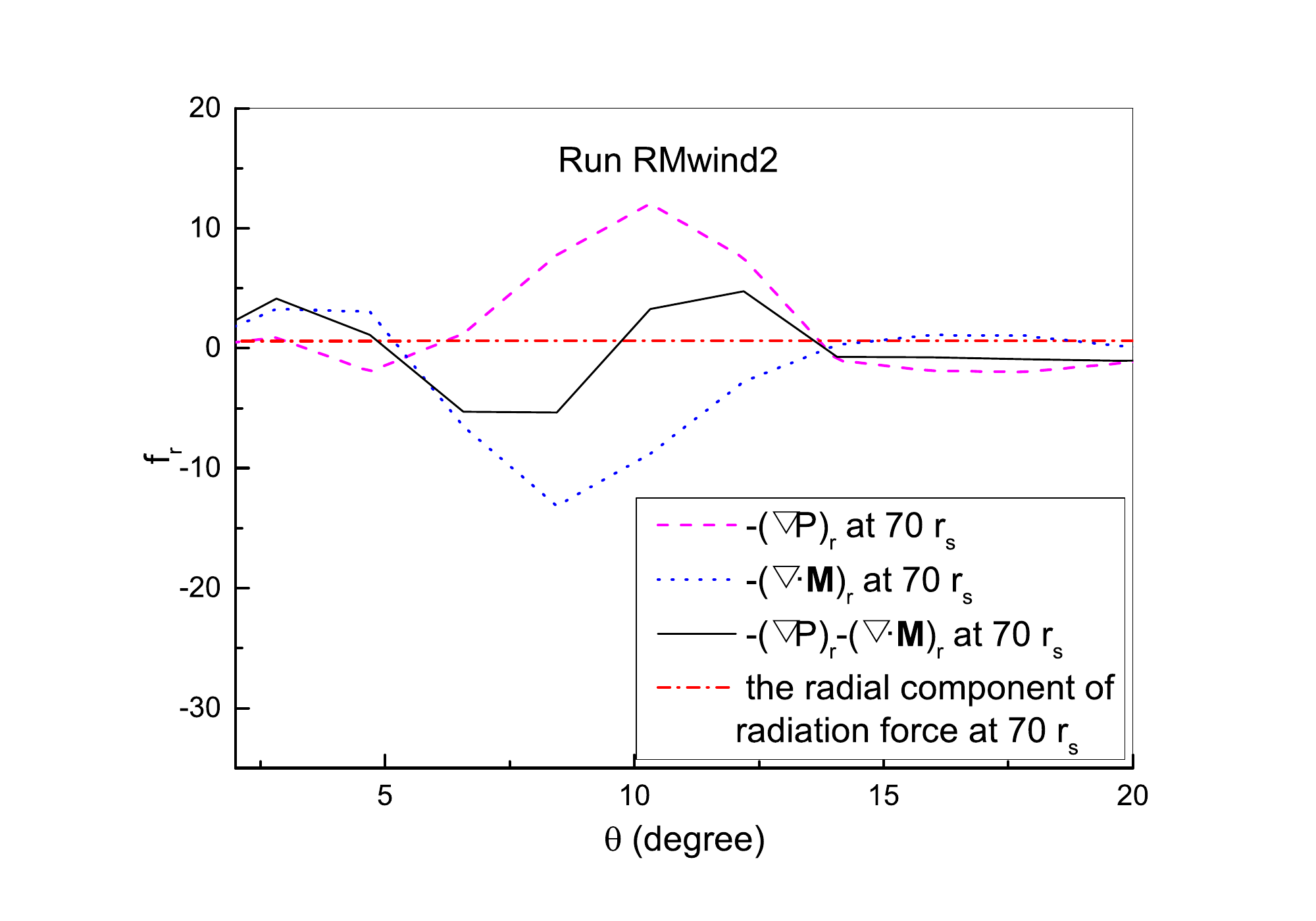}}}
\scalebox{0.28}[0.28]{\rotatebox{0}{\includegraphics[bb=58 33 500 360]{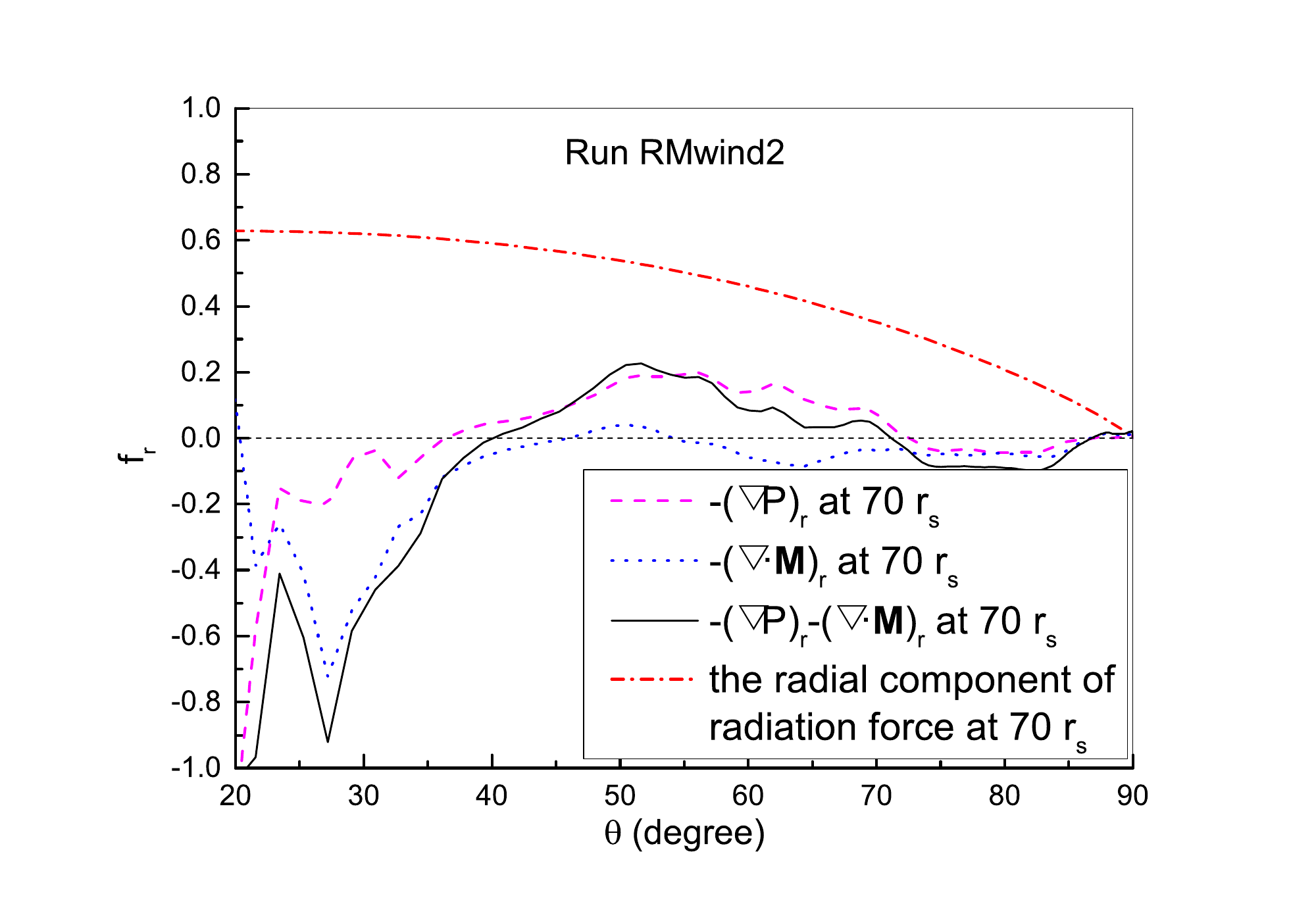}}}

\scalebox{0.28}[0.28]{\rotatebox{0}{\includegraphics[bb=58 33 500 360]{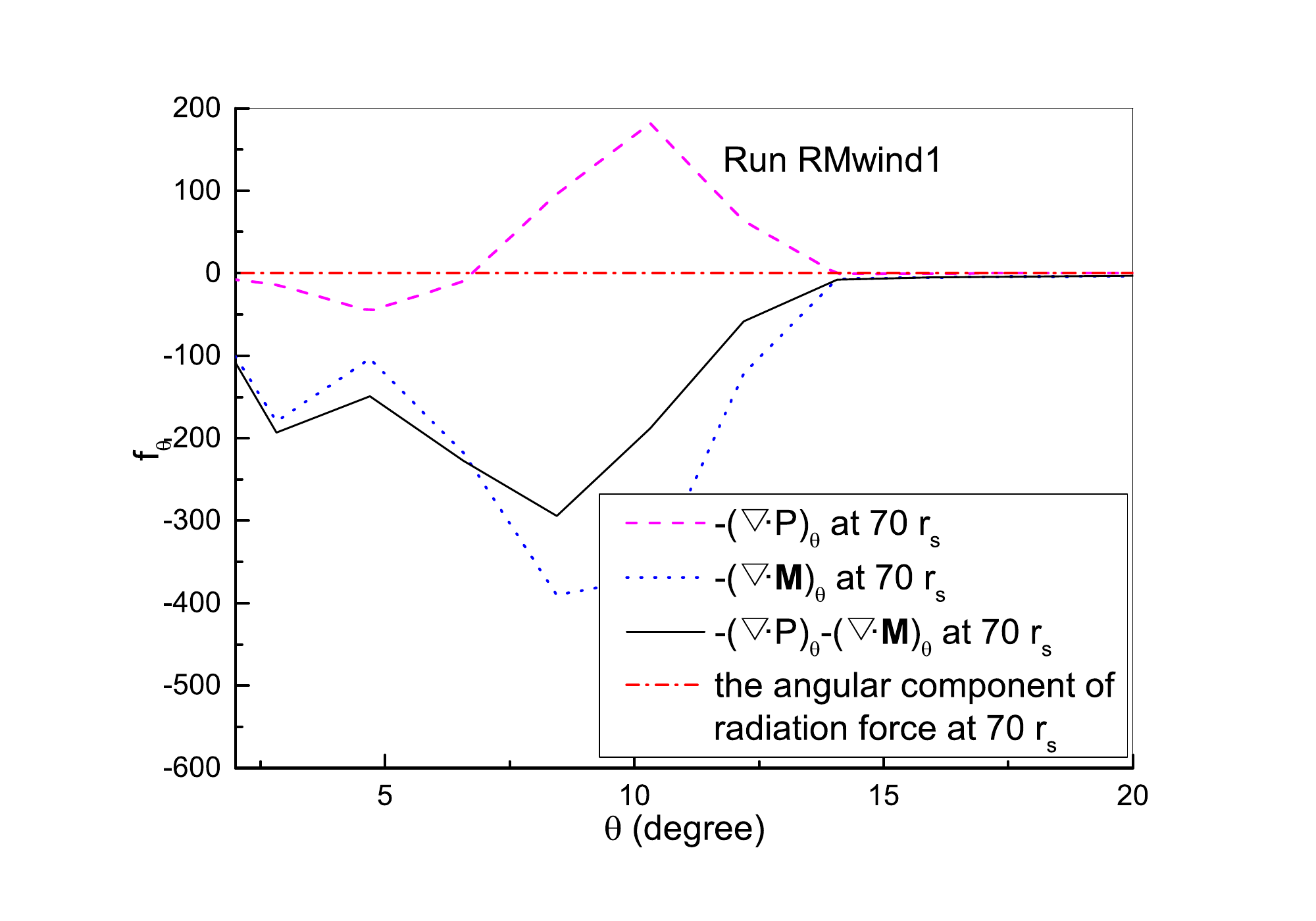}}}
\scalebox{0.28}[0.28]{\rotatebox{0}{\includegraphics[bb=58 33 500 360]{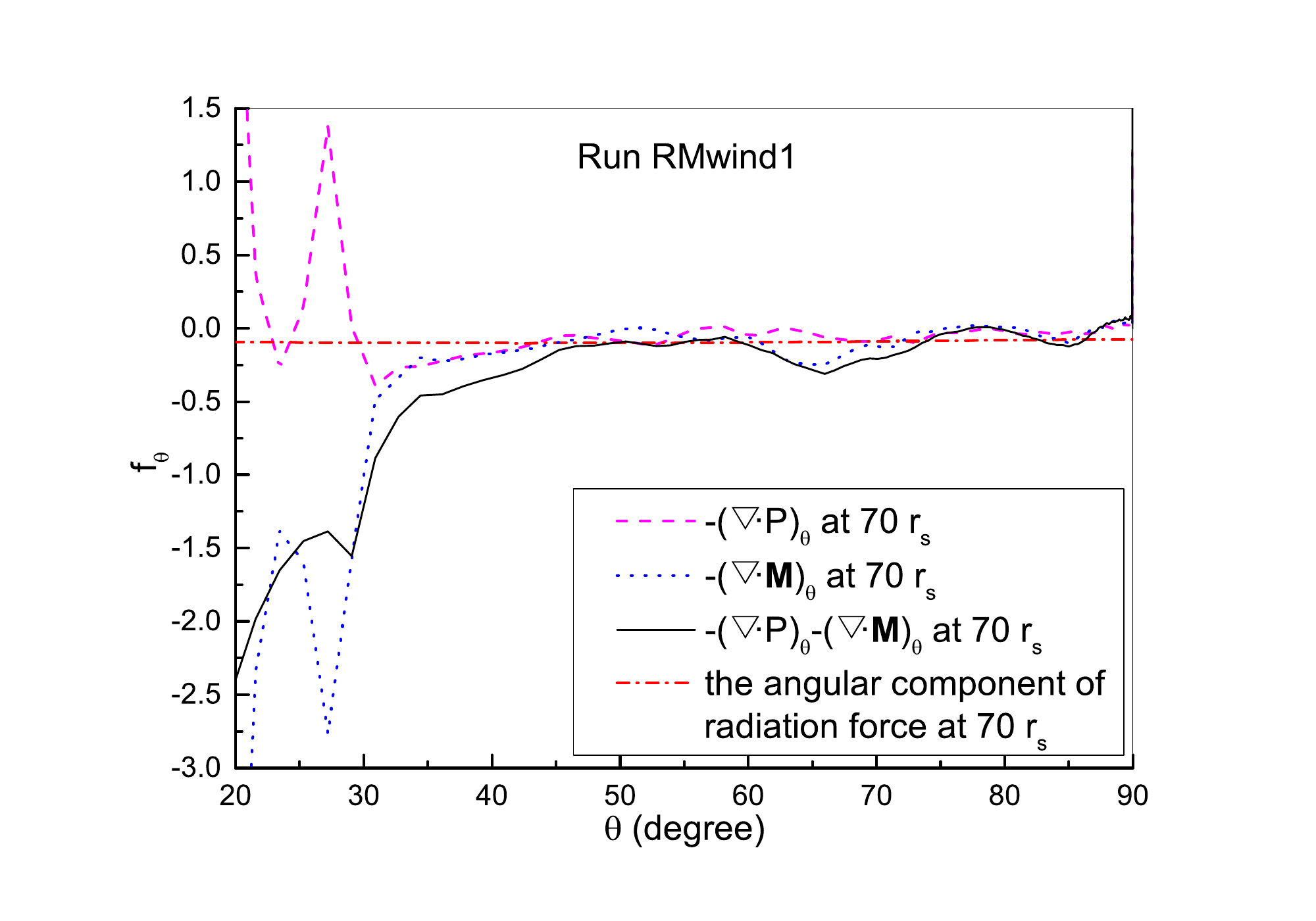}}}
\scalebox{0.28}[0.28]{\rotatebox{0}{\includegraphics[bb=58 33 500 360]{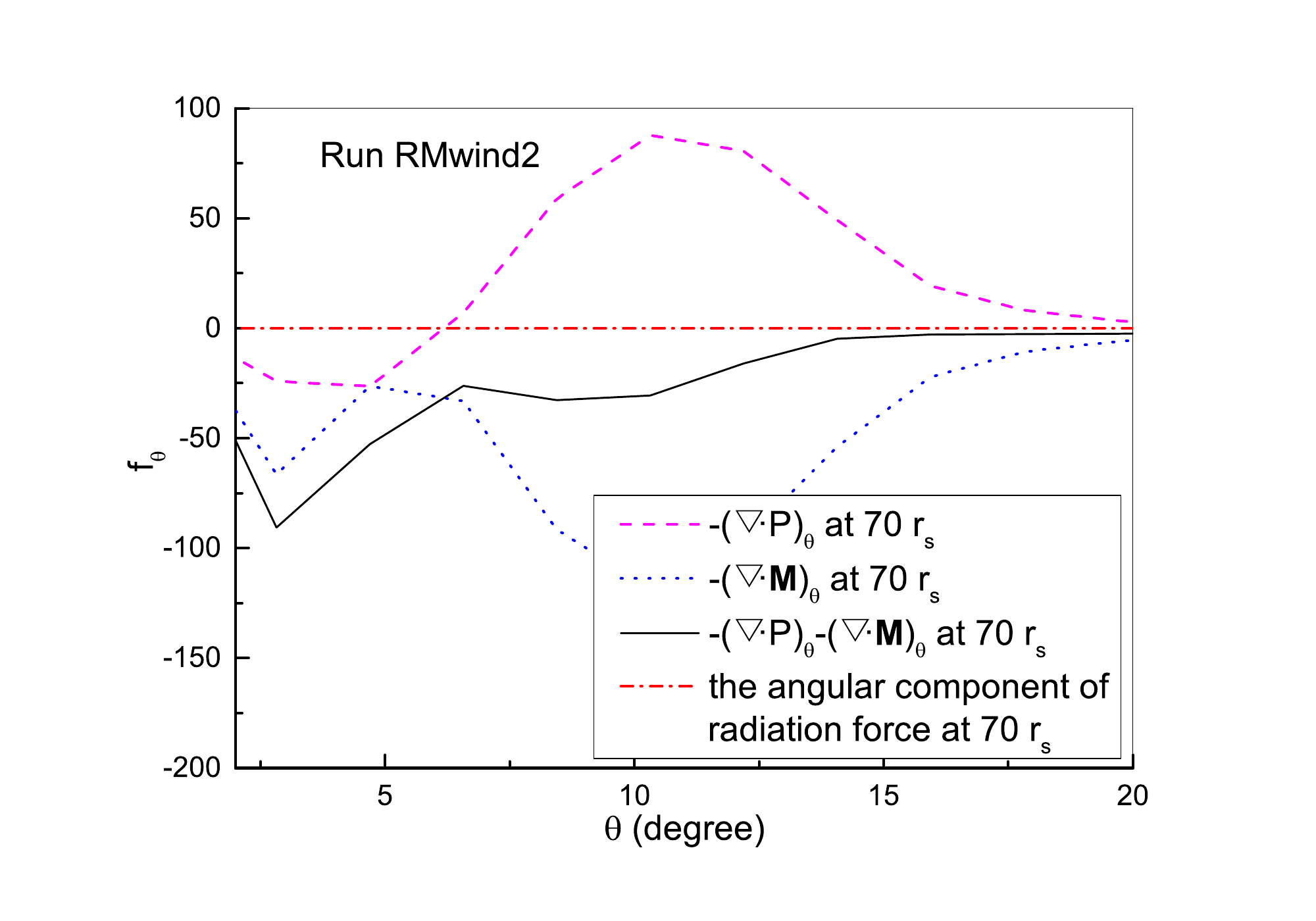}}}
\scalebox{0.28}[0.28]{\rotatebox{0}{\includegraphics[bb=58 33 500 360]{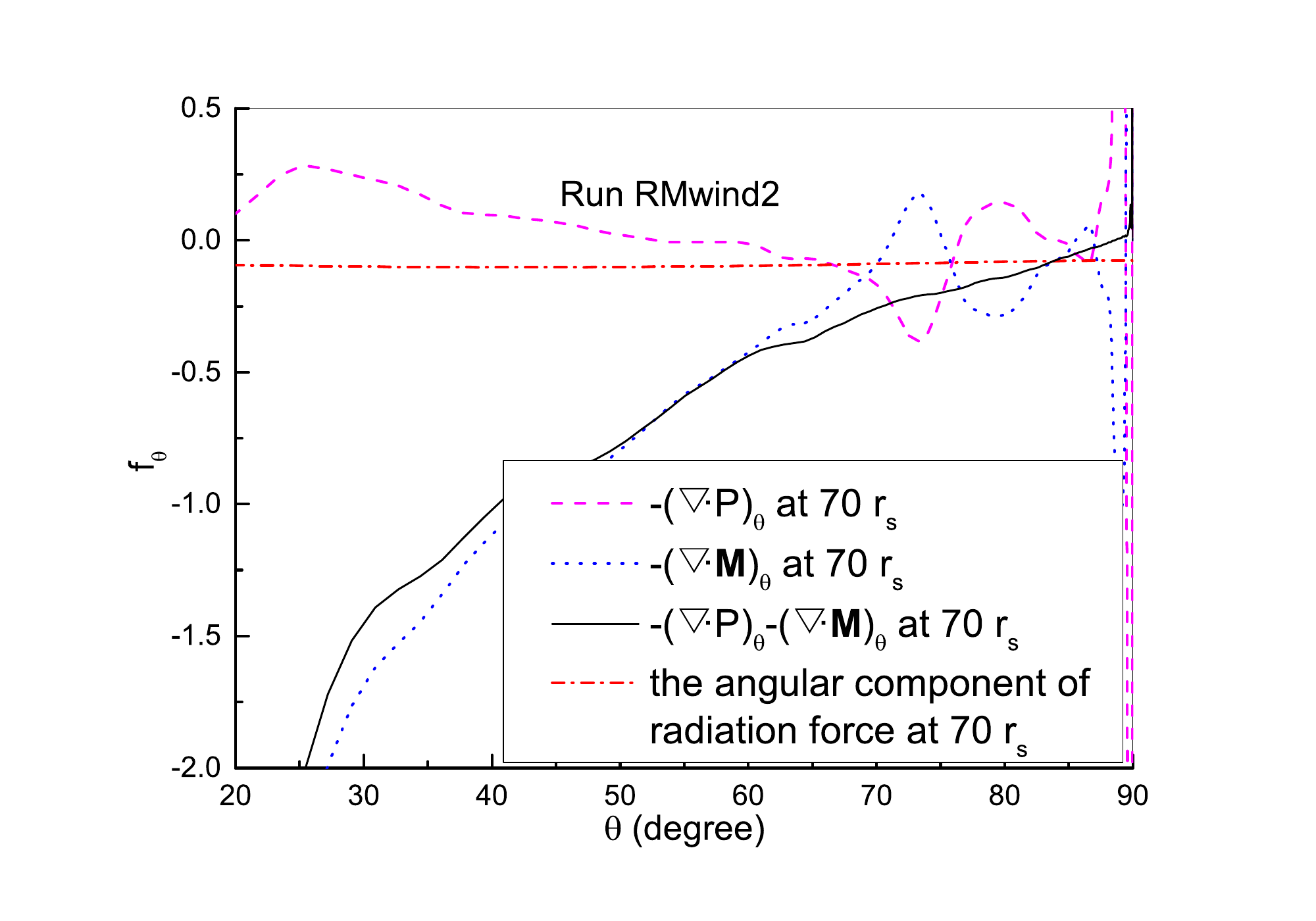}}}

\ \centering \caption{Time-averaged angular distribution of a series of forces in units of the BH Newtonian gravity  ($GM_{\rm bh}/r^2$) at $r=70 r_{\rm s}$. Top panels show the radial component of forces. Bottom panels show the angular component of forces. Dashed lines mean the gradient force ($-\nabla P$) of gas pressure. Dotted lines mean $-\nabla\cdot \bf{M}$, where $\bf{M}$ is the magnetic stress tensor. Dotted-dashed lines mean the radiation force from the thin disk around a BH. Solid lines mean $-\nabla P-\nabla\cdot \bf{M}$. }
\label{fig_8}
\end{figure*}

We plot Figure \ref{fig_6} to quantitatively show the angular dependence of outflows at the outer boundary. To obtain time-averaged values of the variables shown in Figure \ref{fig_6}, we time-average 100 output files over the time range of 2.0--3.0 $T_{\rm orb}$. The variables are the radial column density, the radial velocity, and the mass flux density at the outer boundary, respectively. As shown in Figure \ref{fig_6}, the distribution of radial column density significantly deviates from the initial distribution. Gas expands toward the high latitude. In models with magnetic field, the radial velocity of outflows is higher than $10^4$ km s$^{-1}$ in the region $\theta<10^{o}$ and the maximum value reaches  $\sim10^5$ km s$^{-1}$. The high-velocity outflows are highly collimated (see Figures \ref{fig_2} and \ref{fig_3}) although their mass flux is very low. For the case of low luminosity, such as run RMwind1, turbulence takes place within the angular range of $\theta>\sim30^o$, as shown in Figure \ref{fig_2}, and therefore inflows and outflows are intermittently observed at the outer boundary. In this model, the ordered outflows in the region $\theta<30^{o}$ move outwards with a radial velocity of $\sim10^2$ --$\sim10^5$ km s$^{-1}$. The ordered outflows contribute about 4\% of total mass outflow rate. In model Rwind1, the gas flows inwards in the region of $\theta<80^{o}$. For the case of high luminosity, such as run RMwind2, the outflows at $\theta<60^{o}$ move outwards with a radial velocity of $\sim10^4$ km s$^{-1}$. In model RMwind2, the outflows in the region $\theta<60^{o}$ contribute about 27\% of total mass outflow rate, which is slightly higher than that in the case without magnetic field (Rwind2). The outflows in the case with magnetic field have different properties from those of outflows in models without magnetic field. Important differences are that (1) in the model with magnetic field, the collimated outflows around rotational axis are observed and they have higher radial velocity and (2) in model RMwind1, the middle- and high-latitude outflows have significant contribution to the total outflow mass rate.

The properties of outflows also vary with radius. In order to quantitatively study the radial dependence of outflow properties, we calculate the radial dependence of the mass outflow rate ($\dot {M}_{\rm out} (r)$) and the kinetic ($P_{\rm k} (r)$) and thermal energy ($P_{\rm th} (r)$) carried by the outflow. $\dot {M}_{\rm out} (r)$, $P_{\rm k} (r)$, and $P_{\rm th} (r)$ are, respectively, given by
\begin{equation}
\dot {M}_{\rm out} (r)=4\pi r^2 \int_{\rm 0^\circ}^{\rm 90^\circ}
\rho \max (v_r, 0) \sin\theta d\theta,
\end{equation}

\begin{equation}
P_{\rm k} (r)=2\pi r^2 \int_{\rm 0^\circ}^{\rm 90^\circ} \rho
\max(v_r^3,0) \sin\theta d\theta,
\end{equation}
and
\begin{equation}
P_{\rm th} (r)=4\pi r^2 \int_{\rm 0^\circ}^{\rm 90^\circ} e
\max(v_r,0) \sin\theta d\theta.
\end{equation}
These quantities are time-averaged over $t=2.0$--3.0 $T_{\rm orb}$. Figure \ref{fig_7} presents the time-averaged values of $\dot {M}_{\rm out} (r)$, $P_{\rm k} (r)$ and $P_{\rm th} (r)$ of runs RMwind1 and RMwind2. As shown in panel (A), the mass outflow rate increases outwards for the two models. We use $\dot{M}_{\rm out}(r)\propto r^{\alpha}$ to describe the radial distribution of the mass outflow rate outside 10 $r_{\rm s}$. The values of $\alpha$ are 0.9 and 0.6 for runs RMwind1 and RMwind2, respectively. The averaged value of $\alpha$ approximately equals 0.75. The mass flux at any radius includes the contribution of both the outflows produced locally and the outflows from the smaller radii. The outward-increasing of mass outflow rate implies that outflows can be produced at any radius. The kinetic energy and thermal energy carried by the outflows rapidly increase outwards within $\sim 10 r_{\rm s}$. Outside $\sim 10 r_{\rm s}$, the kinetic and thermal energy fluxes are almost constants with radius.

Bernoulli parameter ($Be$) is an important parameter that is helpful to analyze whether or not outflows can escape from the BH gravitation potential. In the magnetized flows, the Bernoulli parameter is given by (Fukue 1990; Zhu \& Stone 2018)
\begin{equation}
Be\equiv \frac{\textbf{v}^2}{2}+\gamma \frac{P}{(\gamma-1)\rho}-\frac{GM_{\rm BH}}{r-r_{\rm s}}+\frac{B_{\phi}^2}{4\pi\rho}-\frac{B_{\phi}v_{\phi}}{4\pi}\frac{B_{\rm p}}{\rho v_{\rm p}}.
\end{equation}
We have calculated the radial dependence of the Bernoulli parameter of outflows for runs RMwind1 and RMwind2. The Bernoulli parameter is $\theta$-angle-averaged and then time-averaged on the time range of 2.0--3.0 $T_{\rm orb}$. For the collimated outflows, we average the Bernoulli parameter over the angle range of $\theta=0^o$--$10^o$. For the wide-angle ordered outflows, the Bernoulli parameter is averaged over the angle range of $\theta=10^o$--$30^o$ for run RMwind1 and over the angle range of $\theta=10^o$--$60^o$ for run RMwind2. We find that the Bernoulli parameter in both models is positive outside 50 $r_s$, either for the collimated outflows or for the wide-angle ordered outflows. The outflows can escape to infinity.

\subsection{Analysis of the forces exerted on outflows}
The role played by magnetic field on driving outflows from a thin disk was studied analytically in previous works (e.g. Blandford \& Payne 1982; Ferreira 1997). It is found in these previous works that when the poloidal field component is much stronger than the toroidal field component and the energy of the magnetic field dominates the gas energy, gas is enforced by the field line tension to corotate with the disk and then the gas is driven by magnetic-centrifgual force to move outwards. When the toroidal magnetic field is much stronger than the poloidal magnetic field, winds can be driven by magnetic pressure gradient force. To study the role of the Lorentz force in runs RMwind1 and RMwind2, Figure \ref{fig_8} presents time-averaged angular distribution of a series of forces in units of the BH Newtonian gravity ($GM_{\rm BH}/r^2$) at $r=70$ $r_{\rm s}$. In Figure \ref{fig_8}, dashed, dotted, and solid lines represent $-\nabla P$, $-\nabla\cdot \bf{M}$ and $-\nabla P-\nabla\cdot \bf{M}$, respectively, where $\bf{M}$ is the magnetic stress tensor. Dotted-dashed lines represent the radiation force from the thin disk around a BH. Top panels show the radial component and bottom panels show the angular component. As shown in the top panels of Figure \ref{fig_8}, the dominant force driving outflows in the region $\theta<5^{o}$ is $-(\nabla\cdot \bf{M})_{r}$. In the region $\theta>15^{o}$, the radial component of radiation force is much larger than $|(\nabla P+\nabla\cdot \bf{M})_{r}|$. Therefore, the collimated outflows in the region $\theta<5^{o}$ are driven by magnetic field; the outflows in the region $\theta>15^{o}$ are driven by radiation force. In the region $15^o<\theta<45^o$, $-(\nabla\cdot \bf{M})_{r}$ is negative, therefore, $-(\nabla\cdot \bf{M})_{r}$ decelerates the gas in this region. Bottom panels show that $-(\nabla\cdot \bf{M})_{\theta}$ points toward the rotational axis and is a dominating force at the middle and high latitude. This implies that magnetic field can drive gas to move from low latitude to middle and high latitude and enhance outflow mass rate at middle and high latitude, especially in the case of low luminosity. In the angular range of $\theta<15^{o}$, $-\nabla\cdot \bf{M}_{\theta}$ of both models significantly increases and points toward the rotational axis, which is helpful to collimate the outflows around the axis.

\section{Summary}

The hot corona above the thin disk around a BH is irradiated by the thin disk and so the radiation force is exerted on the irradiated corona. A large-scale magnetic field is often considered to be anchored in the thin disk. Therefore, the magnetic field and radiation force are two important factors to drive outflows. We perform two-dimensional magnetohydrodynamical simulations to study the outflows driven by magnetic field and radiation force from the hot corona. We assume that the hot corona is isentropic, i.e., the balance between heating and cooling is kept in the irradiated corona. We employ a weak magnetic field.

Although, an initial weak magnetic field is employed, the magnetic field can significantly change the properties of the outflows driven from the corona. When only radiation force is included in simulations, the collimated outflows around the rotational axis are absent. The middle- and high-latitude outflows disappear in the case of low luminosity. When both the magnetic field and the radiation force are included in our simulations, both the collimated outflows and the wider-angle ordered outflows are observed. The outflows around the rotational axis are highly collimated and move outwards at the speed of $\sim$0.03--0.3 $c$ at the outer boundary. The collimated outflows have very low mass outflow rate. The collimated outflow is a weak jet. The wide-angle ordered outflows are distributed at the middle- and high-latitudes and move outwards with a velocity of $10^2$--$10^4$ km s$^{-1}$. The $\theta$-angle distribution of outflows and the outflow velocity depend on the disk luminosity. Turbulence takes place around the equator.

The magnetic field can accelerate the collimated outflows and drive gas to move from low latitude to middle and high latitudes. In the case of low luminosity, the mass outflow rate at middle and high latitudes is significantly enhanced due to the effect of magnetic field, compared with the case of only including radiation force. In the case of high luminosity, the increase of mass outflow rate is not significant. The Bernoulli parameter of the wide-angle ordered outflows is larger than zero. The mass outflow rate keeps increasing outwards and can be described by $\dot{M}_{\rm out}\propto r^{0.75}$ outside 10 $r_{\rm s}$. The strength of outflows increases with the disk luminosity.

In this paper, the initial magnetic field is weak in the corona above the thin disk. In the future, we will consider the case of strong magnetic field.

\acknowledgments{We thank the anonymous referee for his/her helpful comments. This work is supported by the Fundamental Research Funds for the Central Universities (No. 106112015CDJXY300005 and 2019CDJDWL0005) and the Natural Science Foundation of China (grant 11847301). D. B. is supported in part by the Natural Science Foundation of China (grants  11773053, 11573051, 11633006, and 11661161012), the Natural Science Foundation of Shanghai (grant 16ZR1442200), and the Key Research Program of Frontier Sciences of CAS (No. QYZDJSSW-SYS008). The simulations were carried out at the Super Computing Platform of Shanghai Astronomical Observatory.}

\end{document}